\documentclass[%
 reprint,
 showpacs,
 preprintnumbers,
 amsmath,amssymb,
 aip,
 jcp,
]{revtex4-1}

\usepackage{amsmath}
\usepackage{graphicx}
\usepackage{dcolumn}
\usepackage{bm}
\usepackage{epsfig,color,xspace,multirow,xr,bbold}
\usepackage[all]{xy}
\usepackage{setspace}
\usepackage{url}

\begin{document}

\title{Folding and insertion thermodynamics of the transmembrane WALP peptide}

\author{Tristan Bereau}
 \email{bereau@mpip-mainz.mpg.de}
\affiliation{Max Planck Institute for Polymer Research, Ackermannweg 10, 55128
  Mainz, Germany}
\author{W. F. Drew Bennett}%
\affiliation{Department of Chemistry, 
University of Waterloo, 200 University Avenue West, Waterloo, Ontario, Canada N2L 3G1 }
\author{Jim Pfaendtner}
\affiliation{Department of Chemical Engineering, University of Washington, Seattle, Washington, United States}
\author{Markus Deserno}
\affiliation{Department of Physics, Carnegie Mellon University, Pittsburgh, PA 15213, United States}
\author{Mikko Karttunen}
\affiliation{Department of Mathematics and Computer Science \& 
Institute for Complex Molecular Systems, Eindhoven University 
of Technology - P.O. Box 513, MetaForum, 5600 MB, Eindhoven, 
The Netherlands}

\date{\today}

\begin{abstract}
  The anchor of most integral membrane proteins consists of one or several
  helices spanning the lipid bilayer.  The WALP peptide, GWW(LA)$_n$(L)WWA, is
  a common model helix to study the fundamentals of protein insertion and
  folding, as well as helix-helix association in the membrane. Its structural
  properties have been illuminated in a large number of experimental and
  simulation studies.  In this combined coarse-grained and atomistic
  simulation study, we probe the thermodynamics of a single WALP peptide,
  focusing on both the insertion across the water-membrane interface, as well
  as folding in both water and a membrane.  The potential of mean force
  characterizing the peptide's insertion into the membrane shows qualitatively
  similar behavior across peptides and three force fields.  However, the
  Martini force field exhibits a pronounced secondary minimum for an adsorbed
  interfacial state, which may even become the global minimum---in contrast to
  both atomistic simulations and the alternative PLUM force field.  Even
  though the two coarse-grained models reproduce the free energy of insertion
  of individual amino acids side chains, they both underestimate its
  corresponding value for the full peptide (as compared with atomistic
  simulations), hinting at cooperative physics beyond the residue level.
  Folding of WALP in the two environments indicates the helix as the most
  stable structure, though with different relative stabilities and
  chain-length dependence.
\end{abstract}

\maketitle


\section{Introduction}
\label{sec:intro}

Transmembrane proteins constitute one of the most important biological
building blocks, enabling communication of material and information between a
cell and its environment, or between different intracellular
compartments.\cite{israelachvili1980physical, eisenberg1984three,
  darnell1990molecular, edidin2003lipids} Despite impressive progress in
determining membrane protein structures,\cite{von2011introduction} aided by
technological advances in fields such as electron
tomography\cite{li2013electron} and femtosecond
crystallography,\cite{boutet2012high} the number of known structures still
lags far behind the case of soluble proteins.  Unfortunately, in the absence
of structures, the options for numerical modeling are limited. This is true
not only because protein structure prediction remains a formidable
computational challenge, both for equilibration and force-field
reasons.\cite{bowman2011taming, dill2012protein, Cino2012h,piana2014assessing}
We also face the additional predicament that a lipid bilayer and its
surroundings constitute a very highly anisotropic environment, where
everything from dielectric constants to lateral stresses varies dramatically
on an {\AA}ngstrom scale, pushing both continuum theory and local
thermodynamics to their limits. It should hence not come as a surprise that
even ostensibly basic questions about structure, location and interaction of
small peptides in bilayers remain difficult to
answer.\cite{lindahl2008membrane}

The overwhelming majority of integral membrane proteins is anchored into the
lipid bilayer by one or several transmembrane $\alpha$-helices, followed to a
much smaller fraction by proteins where a $\beta$-barrel motif takes over that
role.\cite{booth1999membrane, bowie2005solving} This is rationalized by the
hydrophobic environment of the lipid tails, which favors protein conformations
that minimize the number of broken backbone hydrogen
bonds.\cite{popot1990membrane, bowie2011membrane}

In an effort to better understand membrane proteins at a biophysical level, a
large body of work has focused on studying individual model helices.  One
common example is the sequence of WALP peptides, composed of alternating
alanine and leucine residues and flanked by two tryptophans at each terminus.
It was designed to resemble a transmembrane helix in membrane proteins, while
permitting an easy way to change its length.\cite{killian1996induction,
  morein1997influence} The arrangement of residues is such that WALP16
corresponds to the sequence GWW(LA)$_5$WWA, while longer WALP peptides include
more LA repeat units (and occasionally an additional alanine between the final
leucine and the C-terminal tryptophans).

Various experimental and simulation studies have shed light on the stability
of WALP as a transmembrane helix. Experimentally, a combination of NMR
methods, hydrogen/deuterium exchange, and mass spectrometry applied to WALP of
different chain lengths, as well as lipids of different size, have provided
important insight into the role of hydrophobic mismatch---the difference
between the length of a peptide's hydrophobic stretch and that of the
bilayer's hydrophobic core.\cite{killian1996induction, morein1997influence}
For instance, a positive mismatch leads to an average tilt angle between the
peptide and the membrane normal, a quantity that can be determined from both
experiments (e.g., quadrupolar splittings from $^2$H solid-state
NMR\cite{strandberg2004tilt}) and computer
simulations.\cite{im2005interfacial, kandasamy2006molecular,
  ozdirekcan2007orientation} Notably, Monticelli \emph{et al.}~resolved an
apparent discrepancy between the average tilt angle extracted from experiment
versus the same observable calculated in molecular dynamics simulations. Using
a coarse-grained model, and hence being able to access much longer time
scales, they showed that both experiment and simulation agree, thus
highlighting the importance of sampling the tilt angle over the microsecond
time scales relevant for NMR experiments.\cite{monticelli2010interpretation}
Atomistic simulations later confirmed these findings using enhanced-sampling
methodologies.\cite{kim2010revisiting}

These studies illuminate the thermodynamics of trans-membrane helices---not
only the stability in the membrane, but also the insertion from water.  Using
an atomistic representation for peptides, but an \emph{implicit}
water/membrane model, Im and Brooks showed that WALP\{16,19,23\}, starting as
an initial random coil, would spontaneously insert and fold into a
bilayer.\cite{im2005interfacial} Further, Nymeyer \emph{et
  al.}\cite{nymeyer2005folding} and Ulmschneider \emph{et
  al.}\cite{ulmschneider2010mechanism} demonstrated insertion and folding of
WALP16 in an \emph{explicit} DPPC membrane using enhanced-sampling
methodologies and high-temperature simulations, respectively, to alleviate the
considerable sampling issues.  Some of us reported similar findings using
PLUM, a recently-developed CG model,\cite{bereau2014more} with and without
enhanced sampling.\cite{bereau2014enhanced}

The potential of mean force (PMF) for the insertion of WALP across a
water/membrane interface provides insight into the thermodynamics of
insertion: both in terms of the free-energy difference between the two
environments and the possible existence of intermediate
barriers. Structurally, WALP is known to form a helix in the membrane, but its
conformation in water is largely unknown, because its many hydrophobic
residues render it prone to aggregation at experimentally relevant
concentrations. Insertion simulations, on the other hand, typically work with
a single peptide (due to sampling limitations). However, their ability to
predict WALP structures in solution is not merely a matter of the required
computational resources, but also of the model's ability to \emph{describe}
secondary structure changes in the first place. For instance, Bond \emph{et
  al.} used CG simulations to study the thermodynamics of insertion of WALP
into a DPPC bilayer. Their model, a variant of the CG Martini force
field,\cite{marrink2007martini} required them to constrain the peptide into a
helix in all environments,\cite{bond2008coarse} which begs the question
whether a potential folding/unfolding equilibrium contributes to the free
energy of insertion.  One aim of our present study is to address this
question.

The following work investigates the link between WALP's structure and its
environment. We rely on the CG PLUM force field\cite{bereau2014more} to
efficiently sample the thermodynamics of insertion across the water-membrane
interface, without explicit bias on the secondary structure.  To gauge the
robustness of the results, we carry out equivalent simulations using both the
CG Martini force field\cite{marrink2007martini}, bearing in mind its
secondary-structure constraints, as well as atomistic simulations, despite
unavoidable challenges associated with sampling.  While the results agree in
many qualitative features, we find a number of interesting exceptions which we
analyze in some detail.  In addition, the free-energy profile as a function of
helicity in both the membrane and water environments provide insight into the
preferred conformations.

\section{Simulation models}
\label{sec:methods}

\subsection{Coarse-grained simulations: PLUM force field}

The following describes the CG PLUM force field. The associated
simulation protocol and parameters used in this work are described
in Appendix~\ref{app:plum}.

The PLUM force field is constructed from the cross-parametrization
of implicit-solvent CG peptide \cite{bereau2009generic} and lipid 
\cite{wang2010systematically, wang2010systematic} models, which we 
summarize in the following.

The peptide model includes amino-acid specificity and can stabilize different
secondary structures using a single parametrization, i.e., without explicit
bias toward one particular conformation.  Each amino acid is described using
four beads: one for the side chain and three for the backbone, providing
enough resolution to describe backbone dihedrals.  Phenomenological
interactions allow the model to reproduce basic properties of peptides and
proteins, such as excluded volume, hydrophobicity, and hydrogen bonds.  The
model was tuned to qualitatively reproduce the Ramachandran plot of
tripeptides and fold a \emph{de novo} three-helix bundle.  Without changing
the force-field parameters, the model can also stabilize different helical
peptides and assemble $\beta$-sheet-rich oligomers.\cite{bereau2009generic}
The CG model has been applied to a variety of scenarios involving helical
peptides,\cite{bereau2010interplay, bereau2011structural} aggregation of
$\beta$-rich peptides,\cite{osborne2013thermodynamic} and $\beta$-barrel
formation at the interface between virus capsid
proteins.\cite{bereau2012coarse}

The lipid model maps a 1-palmitoyl-2-oleoyl-sn-gly-cero-3-phosphocholine
(POPC) lipid into 16 beads, using 8 bead types to distinguish different
chemical moieties.\cite{wang2010systematically} Interaction potentials were
determined from an iterative-Boltzmann inversion\cite{reith2003deriving} of
the radial distribution functions, obtained from an all-atom POPC membrane
simulation. Being an implicit solvent model, the absence of water is
compensated by a phenomenological attractive interaction between tail
beads. Free lipids self-assemble into a bilayer, which then reproduces elastic
properties (e.g., the bending modulus), the mass density profile, and the
orientation of intramolecular bonds.\cite{wang2010systematically} Other
neutral lipids can be constructed from the set of bead types and reach
satisfying transferability in terms of structure, area per lipid, and
temperature dependence of the main phase transition.\cite{wang2010systematic}

While keeping the individual force-field parameters fixed, the
cross-parameters between the peptide and lipid beads were optimized to
reproduce atomistic potential of mean force (PMF) curves of the insertion of
single amino-acid side chains into a DOPC
bilayer.\cite{maccallum2008distribution} The cross-parametrization was
validated by investigating a number of structural properties specific to
membrane peptides, such as tilt angle, hydrophobic mismatch, and transient
pore formation from the cooperative action of antimicrobial
peptides.\cite{bereau2014more} More recently, the use of a Hamiltonian replica
exchange algorithm (more below) assisted in folding several peptides inside
the membrane: WALP\{16,19,23\}, as well as the 50-residue-long major pVIII
coat protein (fd coat) of the filamentous fd
bacteriophage.\cite{bereau2014enhanced}

\subsection{Coarse-grained simulations: Martini force field}

Though Martini is a commonly used force field for the description of
peptide-lipid interactions, we highlight some of the key differences 
with PLUM for completeness. The simulation details used
throughout this work can be found in Appendix~\ref{app:martini}.

The coarse-grained Martini model maps on average four non-hydrogen atoms into
one CG bead, and it can describe a wide variety of biomolecules, e.g., water,
lipids, proteins, carbohydrates, or small molecules.\cite{marrink2007martini,
  de2012improved, monticelli2008martini,lopez2009martini,bereau2015automated}
The key idea is to represent characteristic chemical moyeties with a limited
set of CG bead types, determined from the overall charge, hydrogen-bond
capability, and water/oil partitioning coefficient.\cite{marrink2007martini}

Martini reproduces a number of lipid-membrane characteristics: self-assembly,
area per lipid, elastic properties, as well as a reasonable bilayer stress
profile.\cite{marrink2007martini} A particularly attractive feature of the
model is that its building-block approach eases the construction of a large
variety of molecules, in particular many lipids\cite{ingolfsson2014lipid} and
sterols.\cite{marrink2007martini} Due to the mapping of $3-4$ heavy atoms to 1
bead there can be some ambiguity with regards to the optimal mapping of
molecular fragments. For POPC, the oleyl tail was originally modeled with 5
beads,\cite{marrink2007martini} while an updated model uses 4
beads.\cite{wassenaar2015computational}

Martini has been extended to proteins, focusing mainly on peptide-bilayer
interactions.\cite{monticelli2008martini, de2012improved} The parametrization
quite accurately captures the free-energy of the insertion of single amino
acid side chains and reproduces a number of structural properties of model
transmembrane helices.  Though Martini tends to map a similar number of beads
per amino acid as PLUM, the emphasis is different: a single bead represents
the backbone while several beads constitute each side chain, providing a
better description of the sterics. The single-backbone bead description
necessitates the use of secondary-structure restraints, present in the form of
torsional parameters specific to different folds (e.g., $\alpha$-helix or
$\beta$-sheet). As a result, peptides modeled with Martini cannot (un)fold or
refold during the simulation.

\subsection{Atomistic simulations}

The simulation protocol of the atomistic simulations is detailed in Appendix
\ref{app:aa}.

\section{Results}
\label{sec:results}

\subsection{Insertion thermodynamics}
\label{sec:insert}

\begin{figure*}[htbp]
  \begin{center}
    \includegraphics[width=0.9\linewidth]{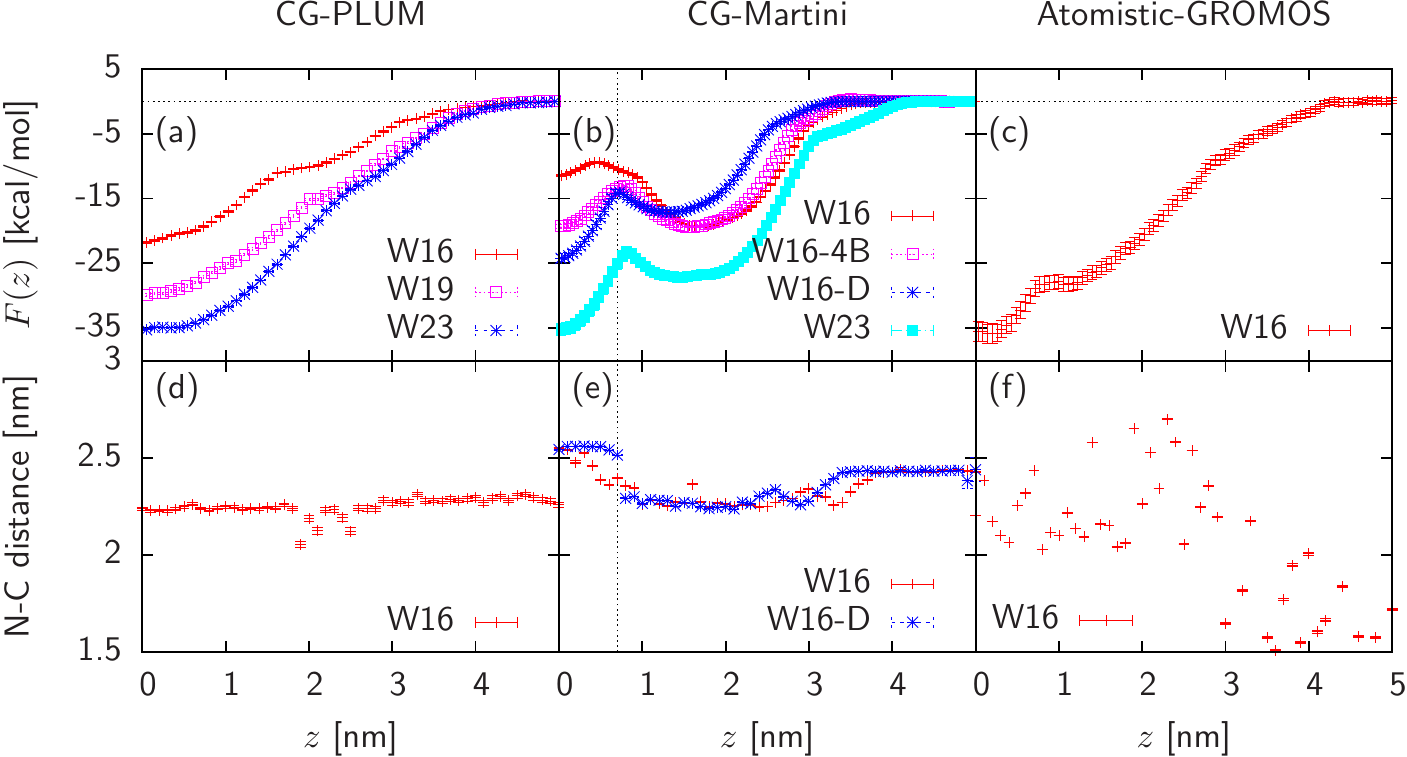}
    \caption{Potential of mean force curves as a function of the distance from
      the bilayer midplane, $z$: (a) WALP\{16,19,23\} from the PLUM force
      field; (b) WALP16 in POPC (``W16''), WALP16 in POPC using the updated
      force field\cite{wassenaar2015computational} with four beads for the
      oleoyl chain (``W16-4B''), WALP16 in DMPC (``W16-D''), and WALP23 in
      POPC (``W23'') using the Martini model with standard water. Note that
      W16 using Martini's polarizable water yielded virtually identical
      results (data not shown); (c) WALP16 using the all-atom GROMOS force
      field.  The PLUM energies are mapped from the reduced unit $\mathcal{E}
      = 0.617$~kcal/mol. (d), (e), and (f): N-C terminus distance (measured
      from C$_\alpha$ to C$_\alpha$) for the respective models. The error of
      the mean is displayed.  Note the larger N-C distance for Martini in DMPC
      (e) close to the bilayer midplane.}
    \label{fig:across}
  \end{center}
\end{figure*}

Fig.~\ref{fig:across} compares the thermodynamics of insertion of a single
WALP\{16,19,23\} peptide into a POPC membrane using different force fields.
In each case, the potential of mean force (PMF) is displayed as a function of
the distance $z$ between the peptide's center of mass and that of the
membrane, which we take as a proxy for its midplane. We extended calculations
up to $z=5$~nm to ensure that the peptide was entirely out of the membrane.

In previous studies, which investigated the insertion of single amino
acids,\cite{maccallum2008distribution, bereau2014more} the bilayer nature was
exploited by simultaneously inserting amino acids into both leaflets. This not
only increased statistics but also minimized conceivable artifacts due to
bilayer asymmetry. In our case the size of a peptide makes this strategy
unfeasible, raising the question how the PMF is affected by this asymmetric
insertion, which stresses the two leaflets differently. In
Appendix~\ref{app:area} we calculate the resulting elastic correction and show
it to be negligible.

\begin{figure}[htbp]
  \begin{center}
    \includegraphics[width=0.9\linewidth]{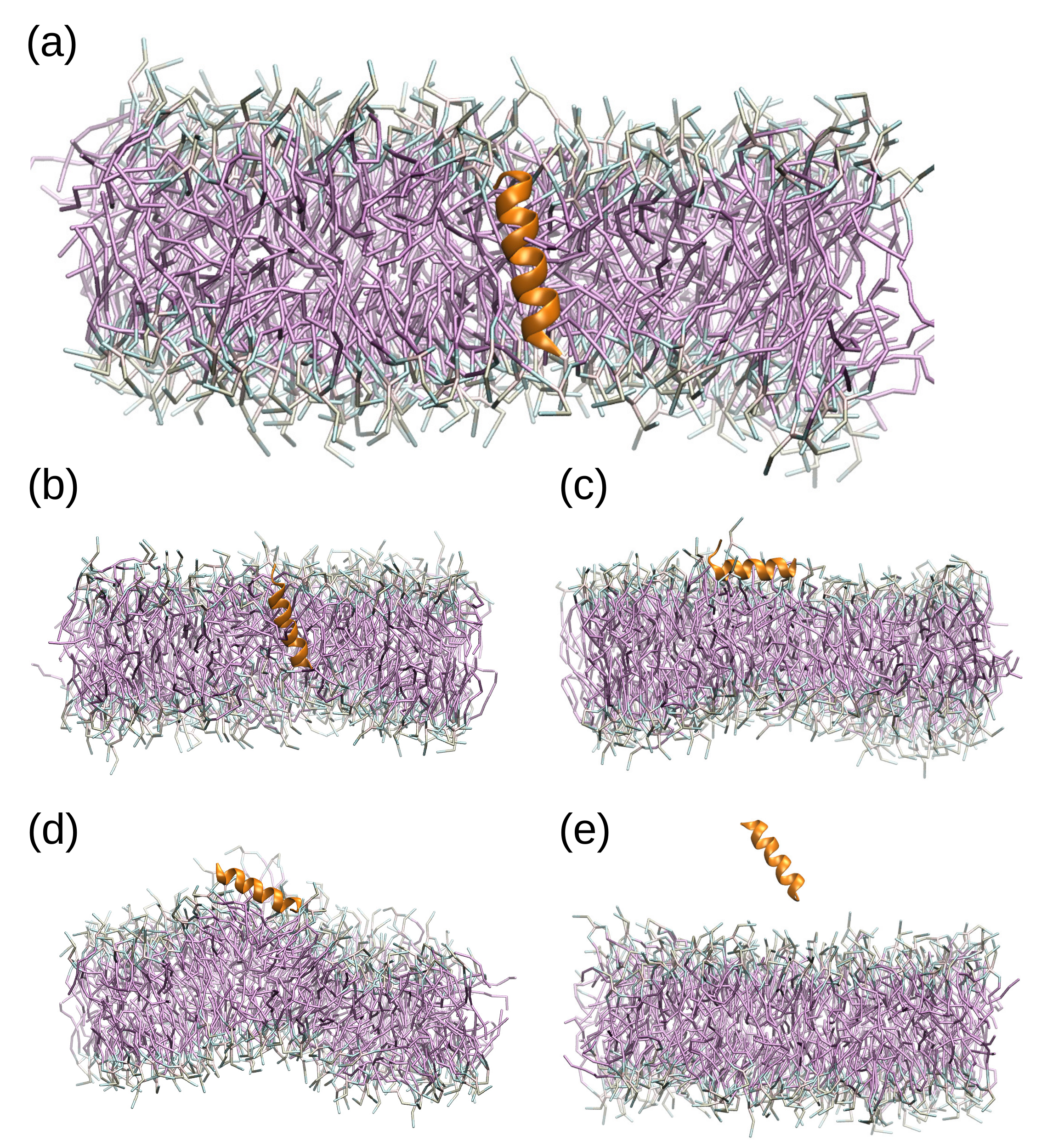}
    \caption{Representative conformations of the WALP16 insertion in POPC
    using the PLUM force field at different distances from the bilayer
    midplane, $z$: (a) $z=0$, (b) $z=1$~nm, (c) $z=2$~nm, (d) $z=3$~nm,
    and (e) $z=4$~nm. The peptide is depicted in orange, where thick 
    and thin ribbons correspond to the helical and coil states, 
    respectively; the lipids are color-coded according to their 
    bead type: purple for the hydrocarbon chains, light pastel colors
    for the interfacial and head groups. Rendered with VMD.\cite{humphrey1996vmd}}
    \label{fig:across_plum}
  \end{center}
\end{figure}

Fig.\ref{fig:across} (a) shows the PMF of WALP16, WALP19, and WALP23 using the
PLUM force field.  All curves indicate that the peptide prefers the bilayer
over the water environment---an expected feature given the hydrophobicity of
the amino acids, and in line with the results of Bond \emph{et
  al.}\cite{bond2008coarse} As we increase the peptide's length, and hence the
number of hydrophobic amino acids, the free energy of the fully inserted state
becomes successively smaller. At the bilayer midplane ($z = 0$), each residue
contributes on average $1.5$~kcal/mol to the free energy of insertion. For
each PMF, we identify three plateaus: ($i$) close to the bilayer midplane
($z\approx 0$) the protein samples transmembrane conformations; ($ii$) around
the bilayer's interfacial region ($z \approx 2$~nm) the peptide is still
helical, but oriented \emph{parallel} to the surface of the membrane; and
finally ($iii$) the asymptotic region ($z \gtrsim 4$~nm) where the peptide has
left the membrane and so its free energy no longer depends on $z$.
Representative conformations are shown in Fig.~\ref{fig:across_plum} for
WALP16, illustrating the transition from fully transmembrane to interfacial to
desorbed.  Notice in particular the significant membrane deformation occurring
at $z \approx 3$~nm (see Fig.~\ref{fig:across_plum} (d)). It occurs because
the peptide's free energy gain for staying in contact with the membrane
outweighs the cost of the concomitant elastic deformation---at least for some
range of $z$-values. Kopelevich recently showed that these deformations lead
to an underestimation of the free-energy barrier upon insertion, though the
overall free-energy difference should be accurate.\cite{kopelevich2013one}

Compared to the PLUM results, the PMFs computed with the Martini model
(Fig.~\ref{fig:across} (b)) have a noticeably different shape. Specifically,
all curves exhibit a secondary minimum corresponding to the interfacial state,
irrespective of whether the standard or the polarizable water model is used
(the two curves overlap, only one of them is shown). While this interfacial
state also exists for the PLUM model, as Fig.~\ref{fig:across_plum} (c)
indicates, its impact on the PMF appears much stronger in the Martini
model. In fact, for WALP16 in a POPC membrane the interfacial state is even
lower in free energy than the completely inserted transmembrane state, and
hence Martini makes a qualitatively different prediction from PLUM about
thermal equilibrium.  In agreement with these results, a spontaneous
transition from transmembrane to interfacial states was previously observed by
Ramadurai \emph{et al.}\cite{ramadurai2010influence}~using unrestrained
simulations of WALP16 in lipid membranes made of five or six tailbead-long
Martini lipids---analogous to the current POPC parametrization.  In fact,
these authors only saw transmembrane-WALP16 spontaneously transition into the
interfacial state when they used lipids with long chains. While they did not
measure a PMF, the barrier from transmembrane to interfacial
(Fig.~\ref{fig:across} (b); $\approx 2$~kcal/mol) calculated by us indeed
suggests the possibility to observe such an event spontaneously, given
reasonably long simulations.

To test whether this behavior originates from the negative hydrophobic
mismatch between WALP16 and POPC, we conducted two control simulations: first,
we kept WALP16 but inserted it into a thinner DMPC bilayer; and second, we
kept the POPC bilayer but used the longer WALP23 peptide.  The resulting PMFs
(Fig.~\ref{fig:across} (b)) show that in both cases the transmembrane state
becomes the most favorable one, even though the interfacial state continues to
produce a very noticeable metastable minimum. This mirrors the observation of
Bond \emph{et al.}, who studied WALP23 in DPPC using a customized version of
the Martini model.\cite{bond2008coarse}

\begin{figure}[htbp]
  \begin{center}
    \includegraphics[width=0.9\linewidth]{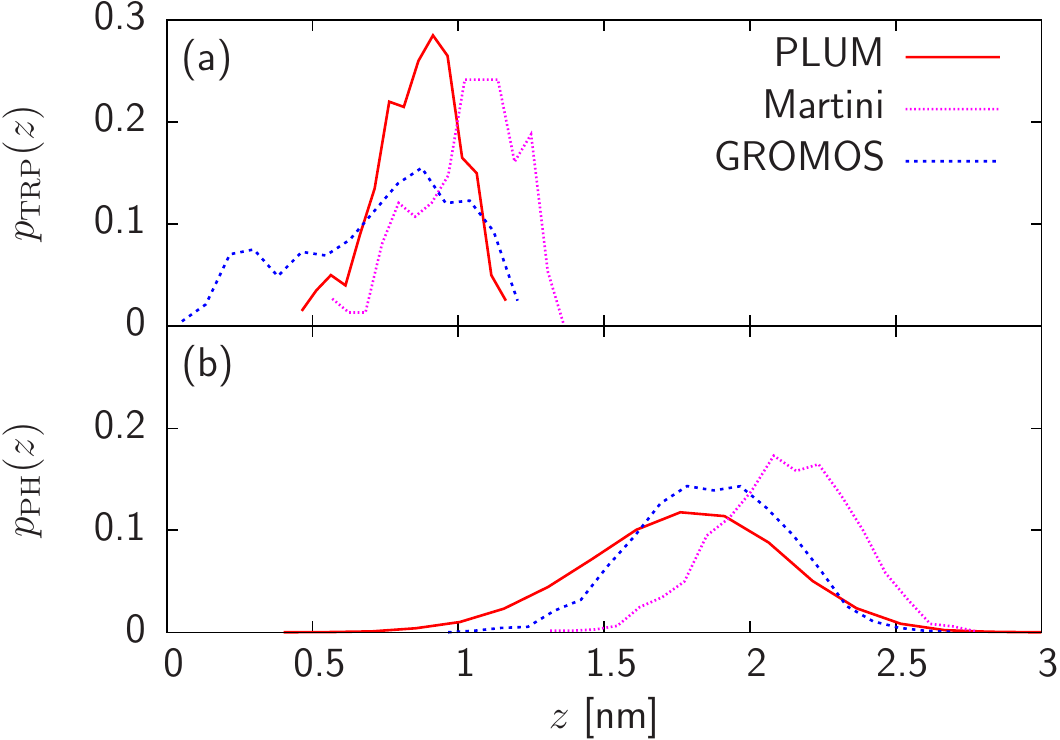}
    \caption{Probability distributions of normal distances between the bilayer
    midplane and (a) the tryptophan side chains, $p_{\rm TRP}(z)$, and (b)
    the lipid phosphate groups, $p_{\rm PH}(z)$, for WALP16 in a POPC
    membrane modeled by PLUM, Martini, and GROMOS. }
    \label{fig:trp_ph}
  \end{center}
\end{figure}

We measured the membrane thickness from the distribution of distances between
the phosphate groups and the bilayer midplane projected along the membrane
normal. While PLUM and GROMOS yield similar distributions that peak around
$1.8$~nm, Martini stabilizes a thicker membrane with a peak around $2.1$~nm
(Fig.~\ref{fig:trp_ph}).  To compare the impact on the alignment of the
peptide, we probed the distribution of distances between the tryptophan side
chains and the bilayer midplane, similarly projected along the membrane
normal. Here again, PLUM and GROMOS yield distributions that peak around the
same point, though the atomistic distribution broadens at lower distances.
Martini, on the other hand, samples a distribution shifted by $\approx 0.1$~nm
to higher values. The differences in offsets between the tryptophan and
phosphate distributions indicate that Martini's thicker membrane will result
in the tryptophan side chains being buried deeper inside the membrane.  Such a
deeper insertion will result in a larger energetic penalty, as evidenced by
the PMF curves of individual side chains\cite{maccallum2008distribution,
  de2012improved, bereau2014more} (sampled using the OPLS force
field,\cite{jorgensen1996development} not GROMOS).

While the current investigation relied on the original POPC Martini model made
of five beads for the oleoyl chain, Wassenaar \emph{et al.}~recently
introduced a parametrization using only four beads, thereby reducing slightly
the membrane thickness.\cite{wassenaar2015computational} The PMF corresponding
to the updated force field is shown in Fig.~\ref{fig:across} (b), ``W16-4B''.
The reduced hydrophobc mismatch between the thinner Martini POPC membrane and
WALP16 lowers the free energy of the transmembrane state, making it roughly
equal to that of the interfacial state. This change goes into the right
direction, but it does not eliminate the pronounced minimum of the interfacial
state, which is absent in the atomistic or PLUM data. This suggests that
hydrophobic mismatch alone is not the sole reason for this feature.

To explore whether the strong hydrophobic mismatch of WALP16 in the 5-bead
Martini POPC membrane also affects the peptide, we monitored the N- to
C-terminal alpha carbon distance as a function of $z$ for all force fields
(Fig.~\ref{fig:across} (d, e, f)) indicates a noticeable stretch for the
Martini peptide in the region $0<z<0.7\,{\rm nm}$, which coincides with the
depth at which the peptide mostly samples a transmembrane helix (data not
shown). WALP16 in POPC shows a gradual decrease of the N-C distance from 2.5
to 2.2~nm between $z=0$ and $z\approx 2$~nm, while WALP16 in DMPC displays a
sudden drop at $z \approx 0.7$~nm, corresponding to the location of the
free-energy barrier in Fig.~\ref{fig:across} (b).  Though Martini stabilizes a
slightly \emph{longer} helix around $z=0$, compared to the other force fields,
the apparent stretching indicates a strong driving force to better accommodate
a short peptide in the bilayer.  On the other hand, PLUM and the atomistic
simulations (described in more details below) do not show any particular
features close to the bilayer midplane.  Unfortunately, the atomistic
N-C-distance data show a lot of scatter, which is clearly a sampling issue.

For WALP16 in DMPC and WALP23 in POPC, the Martini model predicts a pronounced
free energy barrier ($\approx 4$~kcal/mol) for the transition from the
interfacial to the transmembrane state.  This is large enough to become a
problem in unrestrained simulations that aim to study insertion: a peptide
which enters the membrane from the aqueous phase could get trapped in the
interfacial state without transitioning into the transmembrane state, even
though the latter has a free energy that is lower by about $7\,{\rm
  kcal/mol}$. Hall \emph{et al.}\ have indeed encountered this difficulty
during a study that aimed to quantify the insertion thermodynamics of various
WALP peptides in different membranes (using an adapted version of
Martini).\cite{hall2011exploring} They resorted to co-assembling the lipid
bilayer in the presence of a WALP peptide and doing statistics of the final
state thus obtained (inserted or interfacially bound). This protocol suggests
that simply beginning with an interfacially bound peptide was not an option,
for it would rarely if ever proceed to fully insert---a suspicion which the
authors explicitly confirm.

Despite the rather vivid differences in the shape of the PMF, PLUM and Martini
largely agree on the free energy of insertion into the transmembrane state
(meaning, $z=0$), provided the hydrophobic mismatch is relaxed. This is not
completely unexpected, for both models reproduce the PMFs of insertion of
single amino-acid side-chains into a PC bilayer.\cite{bereau2014more,
  monticelli2008martini} The finding is nevertheless nontrivial, because the
absolute values do \emph{not} agree with the atomistic ones, as we will
discuss below.

Fig.~\ref{fig:across} (c) shows the PMF of WALP16 in POPC, using the atomistic
GROMOS force field. $F(z)$ is largely downhill. It exhibits a small shoulder
at $z=1$~nm, but no significant barrier. The location of this shoulder is
close to the point at which the Martini model finally transitions from
interfacial to transmembrane, suggesting that this might indeed be the
physical origin of this feature, but the substantial increase in $F(z)$ by
about $10\,{\rm kcal/mol}$ between $1.7\,{\rm nm}$ and $0.7\,{\rm nm}$
(observed with Martini) is absent. Hence, the general shape of the PMF as
predicted by the PLUM model appears closer to the atomistic data.

Finally, we wish to point out a curious discrepancy between both CG models and
the atomistic reference: in both CG cases the free energy of WALP16 in its
equilibrium state (about $-20\,{\rm kcal/mol}$) is only $2/3$ of the value
predicted in the atomistic simulation (about $-35\,{\rm kcal/mol}$). This is
surprising, because both models capture the free energy of insertion of
individual amino acids, as predicted atomistically. And while especially in
the atomistic case one should always be wary of sampling
issues,\cite{neale2011statistical} and bootstrapping tends to underestimate
error bars, we do not believe that this is the source of the discrepancy, for
it would not suffice to explain a shift by $15\,{\rm kcal/mol}$. Hence, it
seems that the difference is real and has interesting consequences for
modeling. Specifically, it should be clear that the free energy of insertion
of an $\alpha$-helix consisting of $N$ hydrophobic amino acids is not simply
the sum of the free energy of insertion of each individual amino acid, because
there are correlation and cooperativity effects. It seems likely that these
effects depend not just on the physics captured on the coarse-grained level
but on more local effects, too. If so, CG models of peptides will not capture
the insertion free energy correctly, even if ostensibly parametrized for
precisely that, and the difference might even be model dependent. Given the
large amount of research undertaken with these models, it would appear crucial
to understand this issue better.

\subsection{Folding in the membrane}

\begin{figure}[htbp]
  \begin{center}
    \includegraphics[width=0.9\linewidth]{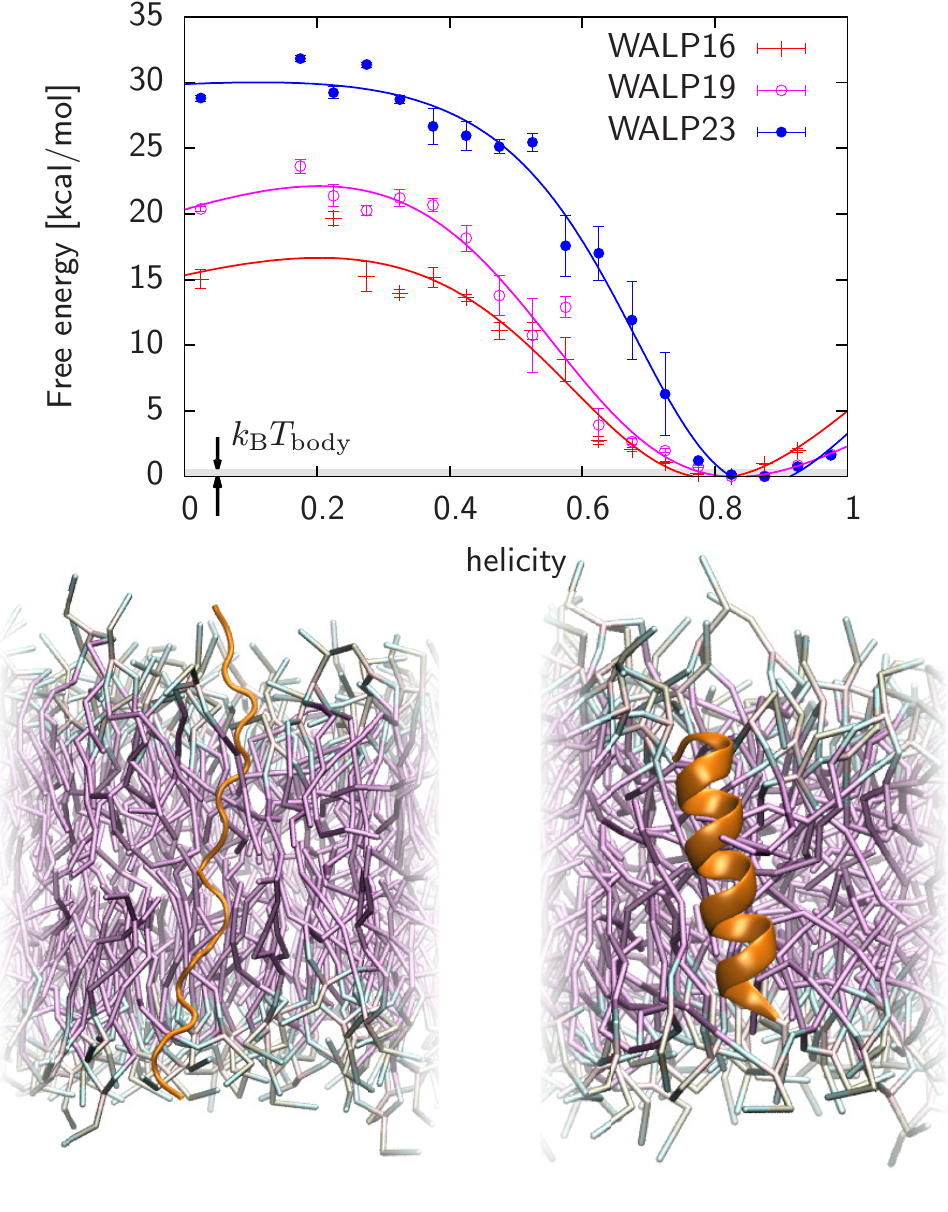}
    \caption{Free energy as a function of helicity for WALP\{16,19,23\}
      inserted in the membrane using the CG PLUM force field.  Solid lines are
      mere guides to the eye.  The grey area roughly indicates the amount of
      thermal energy at body temperature. The conformations underneath
      illustrate an unstructured and helical conformations of WALP16 in the
      membrane. The PLUM energies are mapped from the reduced unit
      $\mathcal{E} = 0.617$~kcal/mol.}
    \label{fig:plum_membrane}
  \end{center}
\end{figure}

Since the PLUM force field was designed to model changes in secondary
structure, we can probe the free-energy landscape of WALP as a function of
helicity---using appropriate techniques to ensure accurate sampling.
Hamiltonian replica exchange molecular dynamics (HREMD) simulations inside the
membrane combined with the weighted histogram analysis method (WHAM; Appendix
\ref{app:WHAM}) yields the free energy profiles shown in
Fig.~\ref{fig:plum_membrane}.  Unsurprisingly, the helical state corresponds
to the free-energy minimum.\cite{kandasamy2006molecular} We note a slight
increase in the free energy when the helicity approaches 1, illustrative of
some fraying at the ends of the chain. The low-helicity states, on the other
hand, are highly suppressed, with free-energy differences ranging from $15$ to
$30$~kcal/mol.  Low but non-zero helicity ($\approx 0.1$) is never observed
due to the secondary-structure prediction algorithm, which relies on the
presence of several ($\approx 4$) consecutive amino acids with appropriate
hydrogen-bonds and dihedrals to assign them in a helical state. We observe a
plateau at low helicity (i.e., $0-0.3$) followed by a sharp,
apparently-downhill profile to the helical state. Overall, we observe a strong
chain-length dependence on the free-energy profile.  If we plot the three
curves against the number of broken backbone hydrogen bonds (data not shown),
the three curves agree more closely in the vicinity of their minima, because
the change in helicity per broken hydrogen bond depends on the peptide's
length.

\subsection{Folding in water}

\begin{figure}[htbp]
  \begin{center}
    \includegraphics[width=0.9\linewidth]{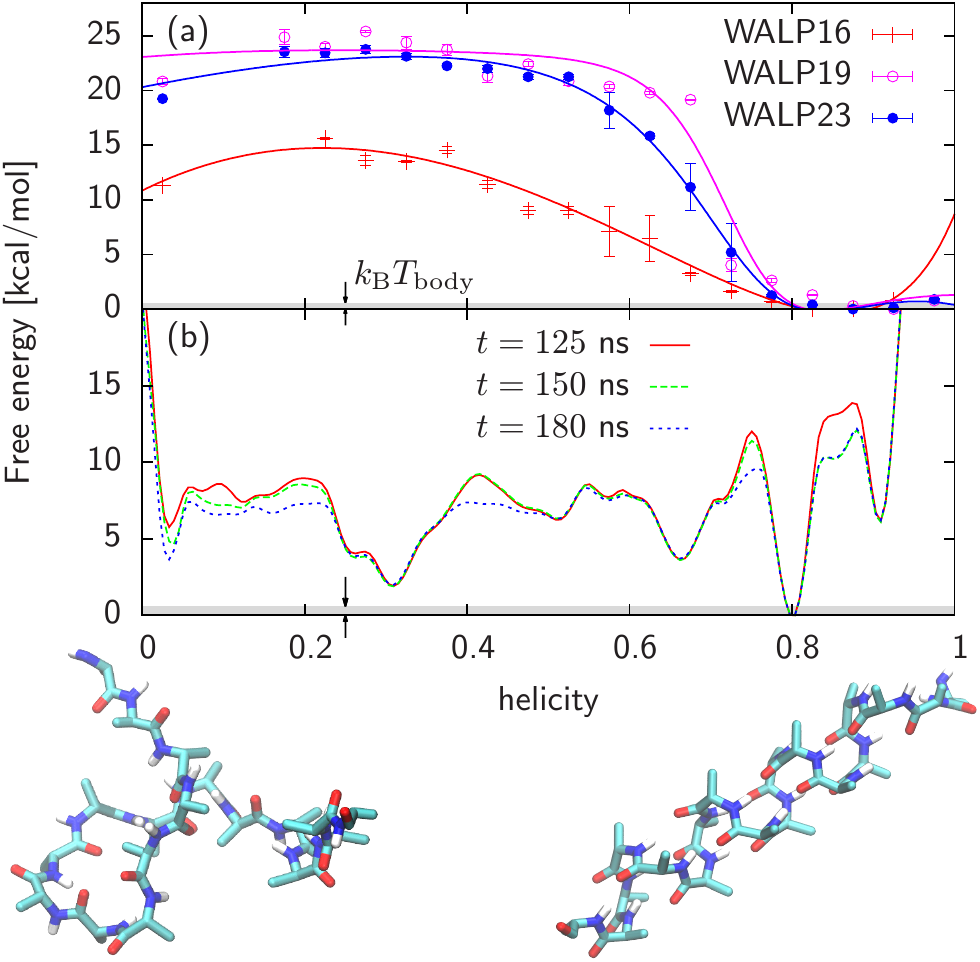}
    \caption{Free energy as a function of helicity for WALP in water using (a)
      CG PLUM simulations on WALP\{16,19,23\} and (b) atomistic simulations on
      WALP16. Solid lines in (a) are mere guides to the eye.  The three curves
      in (b) show the profile at different simulation times per replica, i.e.,
      $t=125$, $150,$ and $180$~ns.  Note the different scales between (a) and
      (b). The grey area roughly indicates the amount of thermal energy at
      body temperature. The CG conformations underneath illustrate random coil
      and helical states of WALP16 in water. The PLUM energies are mapped from
      the reduced unit $\mathcal{E} = 0.617$~kcal/mol.}
    \label{fig:plum_water}
  \end{center}
\end{figure}

We then repeat the free energy study as a function of helicity from the
previous section, but now for WALP dissolved in pure water.
Fig.~\ref{fig:plum_water} (a) shows the free-energy profile of the
WALP\{16,19,23\} peptides simulated using the CG PLUM force field. Just like
in the membrane, we find a strong preference for helical conformations, with
free-energy differences between coil and helix in the range $\Delta F \approx
10-25$~kcal/mol.

Interestingly the chain length dependence of the free-energy is qualitatively
different from the membrane case: while WALP16 again exhibits the lowest free
energy at any value of the helicity, the profiles for WALP19 and WALP23 are
remarkably similar. The qualitative difference between the two environments is
noteworthy, since hydrogen bonds are the most likely contributors to the
free-energy difference. The hydrophobicity will also play a larger role in an
aqueous environment, as compared to the membrane. Yet the interaction strength
of hydrogen bonds in the model does not depend on whether a bead is surrounded
by water or lipids.\cite{bereau2014more} The only noticeable difference
between the formation of a hydrogen bond in water and in the membrane results
from the change between an implicit-water to an explicit-membrane environment,
suggesting an entropic contribution of the model itself. Interestingly, we
also observed a noticeable change in the stability of hydrogen bonds when
transferring a helix from the water to the membrane
environment.\cite{bereau2014more} Overall, this behavior may point at a
complex interplay between the enthalpy (i.e., hydrogen-bonds) and entropy of
helix formation in water,\cite{bereau2010interplay} while hydrophobic residues
immersed in a hydrophobic environment provide more straightforward behavior.

The free-energy profiles shown in Figs.~\ref{fig:plum_membrane} and
\ref{fig:plum_water} (a) are consistent with the insertion process observed in
Sec.~\ref{sec:insert}: the system sampled a majority of helical conformations
both in the fully transmembrane state and in the aqueous region where the
peptide has left the membrane (Fig.~\ref{fig:across_plum} (a) and (e)).

We aimed at comparing these findings against reference atomistic
simulations. Using metadynamics,\cite{laio2002escaping} we computed the
equivalent free-energy profile for WALP16 in water (Fig.~\ref{fig:plum_water}
(b)).  The profile shows a minimum at 80\% helicity, which roughly corresponds
to 10 over the 12 possible hydrogen bonds in the peptide, (indicative of light
fraying of the helix). We observe a fairly complex profile with multiple
minima, all located above the helical state.  The helix is therefore the most
favorable conformation according to these simulations. Compared to the CG
results, we find a much narrower profile around the minimum. This discrepancy
may partially be attributed to the difference in defining hydrogen bonds
between the CG and atomistic simulations (see Appendices \ref{app:plum} and
\ref{app:aa}).  This also impacts the free energy at low helicity: while {\sc
  stride}'s definition of a hydrogen bond does not allow us to observe any
low-helicity conformation (value around 0.1) in the CG profile, the observable
in the metadynamics is continuous along the entire range.

Aside from difficulties to compare the two curves, we point at a possible lack
of sampling: the complexity of the system makes this free-energy profile
difficult to accurately estimate using an atomistic model.  The three curves
shown in Fig.~\ref{fig:plum_water} (b), representing the profile after
simulation times $t=125$, $150,$ and $180$~ns per replica, are illustrative of
the convergence of the profile.  We thus withhold from further interpreting
this curve, and only conclude that the helix may indeed be a relevant
conformation for WALP16 in solution.

\section{Conclusions}
\label{sec:conclusions}

We performed state-of-the-art thermodynamic calculations on WALP peptides
interacting with a model phospholipid membrane using both coarse-grained (CG)
and atomistic force fields.  The potential of mean force (PMF) as a function
of penetration depth $z$ indicates increasing stability as WALP inserts into
the membrane.  PLUM and GROMOS yield qualitatively similar features: an
almost-downhill process from water to the fully-inserted transmembrane
state. Martini, on the other hand, predicts a distinct minimum for the
interfacially bound state, which goes along with a pronounced free-energy
barrier for the transition from the interfacial to the transmembrane state for
all cases we studied. For WALP16 in 5-bead POPC this interfacial minimum even
becomes the global one, in contrast to both PLUM and GROMOS
simulations. Similar behavior was reported in a previous Martini study of
WALP16 as a function of different lipid-tail sizes, where the transmembrane
WALP helix spontaneously flipped to the interfacial state in the presence of
the longer lipids.\cite{ramadurai2010influence} Though the role of negative
hydrophobic mismatch seems to be predominant here, linking this behavior to
particular aspects of the force field remains difficult.

Strikingly, PLUM and Martini report very similar free-energies of insertion at
the bilayer midplane for WALP23 in POPC. The agreement likely results from the
two models' ability to reproduce the insertion of single amino acids in the
bilayer, despite drastically different parametrization
strategies.\cite{monticelli2008martini, bereau2014more} On the other hand, the
atomistic GROMOS simulations suggest increased stability of the transmembrane
helix in the membrane: the atomistic WALP16 curve corresponds roughly to the
insertion of WALP23 in the CG simulations, yielding a discrepancy of $\approx
15$~kcal/mol.  The fact that the two CG models underestimate the free energy
of insertion by the same amount hints at missing correlation and cooperativity
effects beyond the parametrization of individual amino acids.  This poses
questions concerning the coarse-graining strategies for peptides on which both
PLUM and Martini rely, most importantly: under what conditions do matched
thermodynamics on the amino acid level transfer up to the level of a full
peptide? And if it does not, what are the dominant sources of discrepancy and
can we correct for them?

Though Martini enforces secondary structure, making it unable to study the
impact of the environment on folding, PLUM's parametrization did allow us to
probe this behavior in both water and the membrane.  Folding in the membrane
is strongly driven toward the helical state. We find roughly linear
chain-length dependence on the free-energy profile, with longer peptides
increasingly penalizing unstructured random coils.

Folding in water yielded similar behavior as in the membrane, though the
chain-length dependence seems rather different, likely owing to the complex
interplay between secondary structure---hydrogen bonds---and tertiary
structure---hydrophobicity.\cite{bereau2010interplay} The results were
compared with atomistic metadynamics simulations, which also indicate the
helix as the most favorable conformation. Differences in the hydrogen-bond
definition, as well as sampling difficulties of the atomistic model, make it
hard to draw further conclusions. Nevertheless, both models suggest the helix
as a reasonable conformation for WALP in water.

Overall, these findings suggest that enforcing the structure of a helix
throughout the insertion process may reasonably describe the relevant
conformational ensemble of states. In this sense, Martini's lack of peptide
structural rearrangement does not strongly impinge on the results for WALP. A
better understanding of the contribution of (un)folding during the insertion
process will require the study of a different peptide that shows significantly
different folds in water and the membrane. A systematic comparison of such
biomolecular processes using very different computational models (e.g.,
atomistic vs.~coarse-grained or flexible vs.~rigid secondary structure)
provides a better understanding of the impact of their underlying assumption
to large-scale thermodynamic properties.

\begin{acknowledgments}

  We appreciate constructive feedback by Siewert-Jan Marrink on several
  aspects of this work.  We also thank Raffaello Potestio and Joseph
  F.~Rudzinski for a critical reading of the manuscript.  MD acknowledges
  partial support from NSF grant MCB-1330226.  WFDB thanks the Sir Frederick
  Banting Fellowship Program of Natural Sciences and Engineering Research
  Council of Canada for financial support.

\end{acknowledgments}

\appendix
\section{PLUM simulation details}
\label{app:plum}

PLUM's CG units were constructed from a length $\mathcal{L} = 1$~\AA, an energy
$\mathcal{E} = k_{\rm B}T_{\rm body} \approx 0.617$~kcal/mol at $T_{\rm body} = 
310$~K, and a mass $\mathcal{M}$. The time unit $\tau = 
\mathcal{L}\sqrt{\mathcal{M}/\mathcal{E}} \sim 0.1$~ps does not properly 
reflect the dynamics of the system, due to the reduction of 
molecular friction during coarse-graining.\cite{bereau2014enhanced}

We ran all simulations with the {\sc ESPResSo} molecular dynamics
package.\cite{espresso} A Langevin thermostat and modified Andersen barostat
\cite{kolb1999optimized} produced an ensemble with constant temperature ($T=
1.0~\mathcal{E}/k_{\rm B}$), lateral tension ($\Sigma = 0$), and vertical box
height.  Faster integration of the equations of motion was achieved using a
multi-timestepping algorithm, setting the short and long time steps to $\delta
t = 0.01\,\tau$ and $\Delta t = 0.04\,\tau$,
respectively.\cite{bereau2015multi} A 288-POPC lipid membrane was used for all
simulations.  Each peptide was modeled without explicit termini. The helicity
was determined from the {\sc stride} secondary-structure-prediction
algorithm.\cite{stride} More simulation and system-setup details are described
elsewhere.\cite{bereau2014more, bereau2014enhanced}

Hamiltonian replica exchange molecular dynamics (HREMD)
\cite{bunker2000parallel} provided enhanced sampling of a peptide in both
water and the membrane.  In particular, we tuned the strength of the peptide's
hydrogen-bond interaction, i.e., the prefactor $\epsilon$ of a modified
Lennard-Jones potential with added directionality (see Appendix
\ref{app:WHAM}).\cite{bereau2014enhanced} The strength of the interaction was
modulated by a prefactor, $\lambda$, where $\lambda \ge 0$.  We ran HREMD
simulations at prefactor values from $\lambda = 0.1$ to $\lambda = 1.0$,
spanning an appropriate range of conformational space from fully helical to
the complete absence of any helical motif. We used 10 and 20 replicas for
WALP\{16,19\} and WALP23, respectively. Each replica was run for at least
$10^6~\tau$.

To probe insertion thermodynamics, the distance from the bilayer midplane 
to the peptide was measured from the 
$z$-coordinate (i.e., along the bilayer normal) of the center of mass of the 
lipid bilayer to the $z$-coordinate of the center of mass of the peptide. 
Umbrella sampling \cite{Torrie1977} restrained the sampled conformational space
by restraining the normal distance between the $z$-coordinates of the membrane
and the peptide. A harmonic restraint of spring constant $k = 2~\mathcal{E}/{\rm \AA}^2$
was applied at 1~{\AA} intervals, ensuring enough overlap between the different
windows.  In addition, difficulties associated with sampling PMFs of a solute 
in a lipid membrane \cite{neale2011statistical} were addressed here by coupling the
umbrella sampling with HREMD.  Each umbrella restraint was simulated at 4 interaction
prefactors $\lambda \in \{0.55, 0.70, 0.85, 1.00\}$ to help sample the 
conformational flexibility of the peptide. Each replica was run for $10^5~\tau$,
providing an aggregate time of $2\times 10^7~\tau$ for each peptide.

All enhanced-sampling methodologies were unbiased using the weighted histogram
analysis method (WHAM; see also Appendix \ref{app:WHAM}).\cite{FeSw89,
  Kumar92, bereau_swendsen09} All error bars were calculated from
bootstrapping.\cite{resampling}

\section{Martini simulation details}
\label{app:martini}

GROMACS v4.6 \cite{hess2008gromacs} was used for the Martini simulations.
Martini 2.1 \cite{marrink2007martini} and
2.2P\cite{yesylevskyy2010polarizable,de2012improved} models were used for the
standard and polarizable water, respectively.  A 10-fs time step was used,
updating the neighbor list every 10 steps.  Lennard-Jones interactions were
shifted to zero between 0.9 and 1.2 nm.  Electrostatic interactions were
truncated after 1.2 nm with a shifted potential from 0 to 1.2 nm and a
dielectric of 15 (2.5 for polarizable water). Although not recommended to be
used in classical simulations, truncation can be used for Martini due to how
the model has been parametrized.\cite{Cisneros2013} The temperature of 310 K
was maintained using the V-rescale thermostat\cite{bussi2007canonical} with a
1-ps time constant.  Weak semi-isotropic pressure coupling was used with the
Berendsen barostat (1-ps time constant and $3\times 10^{-4}$ bar$^{-1}$
compressibility) \cite{berendsen1984molecular}.  Small bilayer patches were
simulated with 100 POPC lipids per leaflet (121 for the DMPC bilayer). We did
not make use of specific termini groups.

We calculated the free energy for transferring a single WALP from water to the
center of a POPC bilayer (and DMPC for Martini 2.2P).  A cylindrical position
restraint was applied from the C$_\alpha$ of the center residue of WALP and
the center of mass of lipids inside a 1.2-nm-radius cylinder centered around
the peptide, applied along the direction normal to the plane of the
bilayer. To prevent jumps at the cylinder's interface, between 1.2 and 1.7 nm
the weights are switched to zero. A force constant of 500 kJ/mol/nm$^2$ for
the harmonic restraint was used and a 0.1-nm spacing between adjacent umbrella
sampling simulations, from water (5 nm) to the bilayer center (0 nm). Each
simulation was run for at least 500 ns. Free energy profiles were generated
using the weighted histogram analysis method (WHAM) \cite{Kumar92} implemented
in {\sc g\_wham} \cite{hub2010g_wham}.  Error bars were estimated using the
bootstrap method \cite{resampling} with 100 bootstraps.

\section{Atomistic simulation details}
\label{app:aa}

The final WALP16 structure from the Martini umbrella sampling 
simulations was converted back to atomistic representation
using the BACKWARDS \cite{wassenaar2014going} method. 
The GROMOS 54a7 force field \cite{schmid2011definition}
was used on WALP, GROMOS on the POPC lipids,\cite{poger2010new} 
and SPC\cite{berendsen1981interaction} for water. 
We used a 2-fs time step with bonds to hydrogens constrained 
with the LINCS method.\cite{hess1997lincs} The particle mesh Ewald 
summation method 
was used for long-range electrostatic interactions.\cite{essmann1995smooth,Cisneros2013}
Lennard-Jones 
interactions were shifted from 0.9 to 1.0 nm and truncated there after.  
The V-rescale method\cite{bussi2007canonical} was used for temperature 
coupling with a 
reference temperature of 310 K and a 0.1 ps time constant. Pressure 
was maintained semi-isotropically at 1 bar using the Berendsen 
barostat,\cite{berendsen1984molecular} a 2.5-ps time constant 
and $4.5\times 10^{-5}$ bar$^{-1}$ compressibility. 
For the umbrella sampling, we increased the harmonic force constant 
to 3000~kJ/mol/nm$^2$ and ran each simulation for 250~ns.

To compute the free energy of folding in water for WALP16, a combination 
of the metadynamics method and the parallel tempering scheme 
was used.\cite{laio2002escaping, bussi2006free} 
Ten replicas were simulated spanning a temperature range of $290-400$~K. 
An exchange success probability of 9\% was achieved by applying 
the well-tempered ensemble (WTE) approach, which evenly increases 
the spread of the potential energy distribution across replicas, 
while preserving the same ensemble 
averages.\cite{bonomi2010enhanced, deighan2012efficient}
The PTMetaD-WTE calculations were performed and analyzed 
with the PLUMED2 plugin.\cite{tribello2014plumed}

The PTMetaD-WTE simulations were performed for a total 
period of 180 ns/replica. The relative free-energy 
differences between all of the stable minima 
were monitored starting after 100 ns/replica and were 
unchanged after this point, so the simulation was terminated 
after an additional 80 ns/replica. The collective variables, 
$s$, biased in the simulations were a pairwise coordination 
number comprising all of the alpha-helical hydrogen bonds 
($i,i+4$ pairs) and the peptide's radius of gyration (alpha 
carbons only; for a discussion on selecting the collective variables and
a comparison of the different metadynamics techniques, see Ref.~\cite{Do2014}). 
The hydrogen-bond collective variable is formulated in PLUMED 
as a summation of switching functions of the form:
\begin{equation}
s = \sum_{i=1}^{12} 
  \frac{1-\left(\frac{r_{i,i+4}}{r_0}\right)^n}
  {1-\left(\frac{r_{i,i+4}}{r_0}\right)^m}
\end{equation}
with $r_0=0.25$~nm, $n=6$, and $m=9$ (note there are 12 possible 
$\alpha$-helical contacts in WALP16). These numbers 
were scaled by 12 to provide an approximate ``fraction of helicity'' 
in the results.  The metadynamics parameters were 0.2 and 0.01 nm 
for the Gaussian widths of the helicity and radius of gyration.
Gaussians were deposited with a frequency of one per 2 ps, 
and the convergence of the free-energy estimate was 
controlled by using well-tempered metadynamics\cite{barducci2008well} 
and a bias factor of 10. As in previous work, an initial simulation 
period (10 ns/replica) was used to equilibrate the replicas 
to a variety of unfolded structures and build up the WTE energy 
bias to achieve overlap between the 10 replicas. This simulation 
used a bias factor of 40 and a Gaussian width of 450 kJ/mol. 
The results presented show a 1D projection of the 2D metadynamics 
free-energy surface.

\section{Estimating free energies from HREMD using WHAM}
\label{app:WHAM}

The weighted histogram analysis method (WHAM) provides a minimum variance estimator
of the density of states by combining several simulations of the 
same system.\cite{FeSw89, Kumar92} The method is most useful when applied to 
a set of simulations that explore different parts of phase space, each contributing 
to the estimation of thermodynamic properties of the system.  The sampling of phase space
is enhanced by an appropriate choice of Hamiltonians or control parameters (e.g., temperature),
which together help provide a representative sampling of phase space for the process of interest.
Though originally applied to simulations at different temperatures,\cite{FeSw89} 
in the following we vary the Hamiltonian of the original system, $\mathcal{H} = 
\mathcal{H}_0 + V$, where $V$ corresponds to a specific part
of the Hamiltonian, e.g., an interaction potential. 

In this work, we vary the strength of the protein hydrogen-bond 
interaction potential\cite{bereau2009generic, bereau2014enhanced}
\begin{eqnarray}
  V(r,\vartheta_{\rm N},\vartheta_{\rm C}) &=& \epsilon_{\rm hb} \left[ 
    5\left( \frac{\sigma_{\rm hb}}{r}\right)^{12} - 6 \left(\frac{\sigma_{\rm
          hb}}{r}\right)^{10} \right] \\
          && \notag \times \left\{
  \begin{array}{lr}
    \cos^2 \vartheta_{\rm N} \cos^2 \vartheta_{\rm C}, & |\vartheta_{\rm
      N}|,|\vartheta_{\rm C}| < 90^\circ \\
    0 & {\rm otherwise}
  \end{array}\right.
\end{eqnarray}
Each simulation $k$ corresponds to the Hamiltonian $\mathcal{H}_k = \mathcal{H}_0 + \lambda V$,
where $\lambda > 0$. $\lambda=1$ thus corresponds to
the original Hamiltonian, $\mathcal{H}$, while $\lambda \ne 1$ alters the propensity
to form hydrogen bonds.

Assuming that all simulations were run at the same inverse temperature 
$\beta = (k_{\rm B}T)^{-1}$, the calculation of the free energy as a function 
of parameter $\mathcal{Q}$ is provided by
\begin{equation}
\beta F(\mathcal{Q}) \propto -\ln\left[ \sum_{i,s} 
	\frac{\delta(\mathcal{Q}-\mathcal{Q}_{i,s})} 
    {\sum_j N_j \exp\left[\beta(\lambda_i-\lambda_j)V_{s} - f_j\right]} 
    \right],
\end{equation}
where $\delta(*)$ bins parameter $\mathcal{Q}$ in a discrete set, $N_j$ is the
number of samples of simulation $j$, $f_j$ is the scaled free energy of
simulation $j$, $i$ and $j$ sum over simulations, and $s$ sums over
samples.\cite{bereau2011unconstrained} Determination of the set of $f_j$ 
can be obtained by different means.\cite{Kumar92, shirts2008statistically,
bereau_swendsen09, zhu2012convergence}  

\section{Elastic energy of area-leaflet asymmetry upon insertion}
\label{app:area}

Consider a bilayer patch that has an area $A_0$ at zero tension. 
If we insert an object into the upper leaflet that occupies an area $a$, 
the resulting compressive stresses will drive an expansion of that leaflet, 
which in turn puts the lower leaflet under tension. In equilibrium, the bilayer 
expands to an area $A>A_0$, in which a net zero tension arises as a balance of 
compressive and tensile stresses in the upper and lower leaflet, respectively. 
The resulting elastic energy contributes to the free energy of insertion of the 
object. How large is it?

If $K_{A,{\rm m}}=K_A/2$ is the monolayer stretching modulus, the total elastic 
energy can be written as
\begin{equation}
E_{\rm el}= \frac{1}{2}K_{A,{\rm m}}\frac{(A-A_0-a)^2}{A_0} 
+ \frac{1}{2}K_{A,{\rm m}}\frac{(A-A_0)^2}{A_0} \ .
\end{equation}
The still vanishing stress is given by
\begin{equation}
0 = \frac{\partial E_{\rm el}}{\partial A} = 
K_{A,{\rm m}}\left[\frac{A-A_0-a}{A_0}+\frac{A-A_0}{A_0}\right] \ ,
\end{equation}
from which we find $A=A_0+a/2$, showing that the area mismatch is shared evenly 
between the two leaflets. The total elastic energy is therefore
\begin{equation}
E_{\rm el} = \frac{1}{8}K_A A_0 \left(\frac{a}{A_0}\right)^2 \ .
\end{equation}
For WALP, we estimate the area of the inserted object as $a = \pi (d/2)^2$, 
where $d = 12\,{\rm \AA}$ is the diameter of an $\alpha$-helix.  
In our simulations, we use a relaxed membrane area $A_0=(100\,{\rm \AA})^2$, 
such that $(a/A_0)^2 \sim 10^{-4}$. Given a typical value 
$K_A \approx 250\,{\rm mN/m}$ for the stretching modulus,\cite{rawicz2000effect} 
we obtain $E_{\rm el} \sim 0.1\,k_{\rm B}T\sim0.06\,{\rm kcal/mol}$, 
which is a negligible contribution to the overall free energy of insertion.

\bibliography{biblio}

\begin{thebibliography}{84}%
\makeatletter
\providecommand \@ifxundefined [1]{%
 \@ifx{#1\undefined}
}%
\providecommand \@ifnum [1]{%
 \ifnum #1\expandafter \@firstoftwo
 \else \expandafter \@secondoftwo
 \fi
}%
\providecommand \@ifx [1]{%
 \ifx #1\expandafter \@firstoftwo
 \else \expandafter \@secondoftwo
 \fi
}%
\providecommand \natexlab [1]{#1}%
\providecommand \enquote  [1]{``#1''}%
\providecommand \bibnamefont  [1]{#1}%
\providecommand \bibfnamefont [1]{#1}%
\providecommand \citenamefont [1]{#1}%
\providecommand \href@noop [0]{\@secondoftwo}%
\providecommand \href [0]{\begingroup \@sanitize@url \@href}%
\providecommand \@href[1]{\@@startlink{#1}\@@href}%
\providecommand \@@href[1]{\endgroup#1\@@endlink}%
\providecommand \@sanitize@url [0]{\catcode `\\12\catcode `\$12\catcode
  `\&12\catcode `\#12\catcode `\^12\catcode `\_12\catcode `\%12\relax}%
\providecommand \@@startlink[1]{}%
\providecommand \@@endlink[0]{}%
\providecommand \url  [0]{\begingroup\@sanitize@url \@url }%
\providecommand \@url [1]{\endgroup\@href {#1}{\urlprefix }}%
\providecommand \urlprefix  [0]{URL }%
\providecommand \Eprint [0]{\href }%
\providecommand \doibase [0]{http://dx.doi.org/}%
\providecommand \selectlanguage [0]{\@gobble}%
\providecommand \bibinfo  [0]{\@secondoftwo}%
\providecommand \bibfield  [0]{\@secondoftwo}%
\providecommand \translation [1]{[#1]}%
\providecommand \BibitemOpen [0]{}%
\providecommand \bibitemStop [0]{}%
\providecommand \bibitemNoStop [0]{.\EOS\space}%
\providecommand \EOS [0]{\spacefactor3000\relax}%
\providecommand \BibitemShut  [1]{\csname bibitem#1\endcsname}%
\let\auto@bib@innerbib\@empty
\bibitem [{\citenamefont {Israelachvili}, \citenamefont {Mar{\v{c}}elja},\ and\
  \citenamefont {Horn}(1980)}]{israelachvili1980physical}%
  \BibitemOpen
  \bibfield  {author} {\bibinfo {author} {\bibfnamefont {J.}~\bibnamefont
  {Israelachvili}}, \bibinfo {author} {\bibfnamefont {S.}~\bibnamefont
  {Mar{\v{c}}elja}}, \ and\ \bibinfo {author} {\bibfnamefont {R.~G.}\
  \bibnamefont {Horn}},\ }\href@noop {} {\bibfield  {journal} {\bibinfo
  {journal} {Quarterly reviews of biophysics}\ }\textbf {\bibinfo {volume}
  {13}},\ \bibinfo {pages} {121} (\bibinfo {year} {1980})}\BibitemShut
  {NoStop}%
\bibitem [{\citenamefont {Eisenberg}(1984)}]{eisenberg1984three}%
  \BibitemOpen
  \bibfield  {author} {\bibinfo {author} {\bibfnamefont {D.}~\bibnamefont
  {Eisenberg}},\ }\href@noop {} {\bibfield  {journal} {\bibinfo  {journal}
  {Annual review of biochemistry}\ }\textbf {\bibinfo {volume} {53}},\ \bibinfo
  {pages} {595} (\bibinfo {year} {1984})}\BibitemShut {NoStop}%
\bibitem [{\citenamefont {Darnell}\ \emph {et~al.}(1990)\citenamefont
  {Darnell}, \citenamefont {Lodish}, \citenamefont {Baltimore} \emph
  {et~al.}}]{darnell1990molecular}%
  \BibitemOpen
  \bibfield  {author} {\bibinfo {author} {\bibfnamefont {J.~E.}\ \bibnamefont
  {Darnell}}, \bibinfo {author} {\bibfnamefont {H.~F.}\ \bibnamefont {Lodish}},
  \bibinfo {author} {\bibfnamefont {D.}~\bibnamefont {Baltimore}},  \emph
  {et~al.},\ }\href@noop {} {\emph {\bibinfo {title} {Molecular cell
  biology}}},\ Vol.~\bibinfo {volume} {2}\ (\bibinfo  {publisher} {Scientific
  American Books New York},\ \bibinfo {year} {1990})\BibitemShut {NoStop}%
\bibitem [{\citenamefont {Edidin}(2003)}]{edidin2003lipids}%
  \BibitemOpen
  \bibfield  {author} {\bibinfo {author} {\bibfnamefont {M.}~\bibnamefont
  {Edidin}},\ }\href@noop {} {\bibfield  {journal} {\bibinfo  {journal} {Nature
  Reviews Molecular Cell Biology}\ }\textbf {\bibinfo {volume} {4}},\ \bibinfo
  {pages} {414} (\bibinfo {year} {2003})}\BibitemShut {NoStop}%
\bibitem [{\citenamefont {von Heijne}(2011)}]{von2011introduction}%
  \BibitemOpen
  \bibfield  {author} {\bibinfo {author} {\bibfnamefont {G.}~\bibnamefont {von
  Heijne}},\ }\href@noop {} {\bibfield  {journal} {\bibinfo  {journal} {Annual
  review of biochemistry}\ }\textbf {\bibinfo {volume} {80}},\ \bibinfo {pages}
  {157} (\bibinfo {year} {2011})}\BibitemShut {NoStop}%
\bibitem [{\citenamefont {Li}\ \emph {et~al.}(2013)\citenamefont {Li},
  \citenamefont {Mooney}, \citenamefont {Zheng}, \citenamefont {Booth},
  \citenamefont {Braunfeld}, \citenamefont {Gubbens}, \citenamefont {Agard},\
  and\ \citenamefont {Cheng}}]{li2013electron}%
  \BibitemOpen
  \bibfield  {author} {\bibinfo {author} {\bibfnamefont {X.}~\bibnamefont
  {Li}}, \bibinfo {author} {\bibfnamefont {P.}~\bibnamefont {Mooney}}, \bibinfo
  {author} {\bibfnamefont {S.}~\bibnamefont {Zheng}}, \bibinfo {author}
  {\bibfnamefont {C.~R.}\ \bibnamefont {Booth}}, \bibinfo {author}
  {\bibfnamefont {M.~B.}\ \bibnamefont {Braunfeld}}, \bibinfo {author}
  {\bibfnamefont {S.}~\bibnamefont {Gubbens}}, \bibinfo {author} {\bibfnamefont
  {D.~A.}\ \bibnamefont {Agard}}, \ and\ \bibinfo {author} {\bibfnamefont
  {Y.}~\bibnamefont {Cheng}},\ }\href@noop {} {\bibfield  {journal} {\bibinfo
  {journal} {Nature methods}\ }\textbf {\bibinfo {volume} {10}},\ \bibinfo
  {pages} {584} (\bibinfo {year} {2013})}\BibitemShut {NoStop}%
\bibitem [{\citenamefont {Boutet}\ \emph {et~al.}(2012)\citenamefont {Boutet},
  \citenamefont {Lomb}, \citenamefont {Williams}, \citenamefont {Barends},
  \citenamefont {Aquila}, \citenamefont {Doak}, \citenamefont {Weierstall},
  \citenamefont {DePonte}, \citenamefont {Steinbrener}, \citenamefont {Shoeman}
  \emph {et~al.}}]{boutet2012high}%
  \BibitemOpen
  \bibfield  {author} {\bibinfo {author} {\bibfnamefont {S.}~\bibnamefont
  {Boutet}}, \bibinfo {author} {\bibfnamefont {L.}~\bibnamefont {Lomb}},
  \bibinfo {author} {\bibfnamefont {G.~J.}\ \bibnamefont {Williams}}, \bibinfo
  {author} {\bibfnamefont {T.~R.}\ \bibnamefont {Barends}}, \bibinfo {author}
  {\bibfnamefont {A.}~\bibnamefont {Aquila}}, \bibinfo {author} {\bibfnamefont
  {R.~B.}\ \bibnamefont {Doak}}, \bibinfo {author} {\bibfnamefont
  {U.}~\bibnamefont {Weierstall}}, \bibinfo {author} {\bibfnamefont {D.~P.}\
  \bibnamefont {DePonte}}, \bibinfo {author} {\bibfnamefont {J.}~\bibnamefont
  {Steinbrener}}, \bibinfo {author} {\bibfnamefont {R.~L.}\ \bibnamefont
  {Shoeman}},  \emph {et~al.},\ }\href@noop {} {\bibfield  {journal} {\bibinfo
  {journal} {Science}\ }\textbf {\bibinfo {volume} {337}},\ \bibinfo {pages}
  {362} (\bibinfo {year} {2012})}\BibitemShut {NoStop}%
\bibitem [{\citenamefont {Bowman}, \citenamefont {Voelz},\ and\ \citenamefont
  {Pande}(2011)}]{bowman2011taming}%
  \BibitemOpen
  \bibfield  {author} {\bibinfo {author} {\bibfnamefont {G.~R.}\ \bibnamefont
  {Bowman}}, \bibinfo {author} {\bibfnamefont {V.~A.}\ \bibnamefont {Voelz}}, \
  and\ \bibinfo {author} {\bibfnamefont {V.~S.}\ \bibnamefont {Pande}},\
  }\href@noop {} {\bibfield  {journal} {\bibinfo  {journal} {Current opinion in
  structural biology}\ }\textbf {\bibinfo {volume} {21}},\ \bibinfo {pages} {4}
  (\bibinfo {year} {2011})}\BibitemShut {NoStop}%
\bibitem [{\citenamefont {Dill}\ and\ \citenamefont
  {MacCallum}(2012)}]{dill2012protein}%
  \BibitemOpen
  \bibfield  {author} {\bibinfo {author} {\bibfnamefont {K.~A.}\ \bibnamefont
  {Dill}}\ and\ \bibinfo {author} {\bibfnamefont {J.~L.}\ \bibnamefont
  {MacCallum}},\ }\href@noop {} {\bibfield  {journal} {\bibinfo  {journal}
  {Science}\ }\textbf {\bibinfo {volume} {338}},\ \bibinfo {pages} {1042}
  (\bibinfo {year} {2012})}\BibitemShut {NoStop}%
\bibitem [{\citenamefont {Cino}, \citenamefont {Choy},\ and\ \citenamefont
  {Karttunen}(2012)}]{Cino2012h}%
  \BibitemOpen
  \bibfield  {author} {\bibinfo {author} {\bibfnamefont {E.~A.}\ \bibnamefont
  {Cino}}, \bibinfo {author} {\bibfnamefont {W.-Y.}\ \bibnamefont {Choy}}, \
  and\ \bibinfo {author} {\bibfnamefont {M.}~\bibnamefont {Karttunen}},\ }\href
  {\doibase 10.1021/ct300323g} {\bibfield  {journal} {\bibinfo  {journal} {J
  Chem Theory Comput}\ }\textbf {\bibinfo {volume} {8}},\ \bibinfo {pages}
  {2725} (\bibinfo {year} {2012})}\BibitemShut {NoStop}%
\bibitem [{\citenamefont {Piana}, \citenamefont {Klepeis},\ and\ \citenamefont
  {Shaw}(2014)}]{piana2014assessing}%
  \BibitemOpen
  \bibfield  {author} {\bibinfo {author} {\bibfnamefont {S.}~\bibnamefont
  {Piana}}, \bibinfo {author} {\bibfnamefont {J.~L.}\ \bibnamefont {Klepeis}},
  \ and\ \bibinfo {author} {\bibfnamefont {D.~E.}\ \bibnamefont {Shaw}},\
  }\href@noop {} {\bibfield  {journal} {\bibinfo  {journal} {Current opinion in
  structural biology}\ }\textbf {\bibinfo {volume} {24}},\ \bibinfo {pages}
  {98} (\bibinfo {year} {2014})}\BibitemShut {NoStop}%
\bibitem [{\citenamefont {Lindahl}\ and\ \citenamefont
  {Sansom}(2008)}]{lindahl2008membrane}%
  \BibitemOpen
  \bibfield  {author} {\bibinfo {author} {\bibfnamefont {E.}~\bibnamefont
  {Lindahl}}\ and\ \bibinfo {author} {\bibfnamefont {M.~S.}\ \bibnamefont
  {Sansom}},\ }\href@noop {} {\bibfield  {journal} {\bibinfo  {journal}
  {Current opinion in structural biology}\ }\textbf {\bibinfo {volume} {18}},\
  \bibinfo {pages} {425} (\bibinfo {year} {2008})}\BibitemShut {NoStop}%
\bibitem [{\citenamefont {Booth}\ and\ \citenamefont
  {Curran}(1999)}]{booth1999membrane}%
  \BibitemOpen
  \bibfield  {author} {\bibinfo {author} {\bibfnamefont {P.~J.}\ \bibnamefont
  {Booth}}\ and\ \bibinfo {author} {\bibfnamefont {A.~R.}\ \bibnamefont
  {Curran}},\ }\href@noop {} {\bibfield  {journal} {\bibinfo  {journal}
  {Current opinion in structural biology}\ }\textbf {\bibinfo {volume} {9}},\
  \bibinfo {pages} {115} (\bibinfo {year} {1999})}\BibitemShut {NoStop}%
\bibitem [{\citenamefont {Bowie}(2005)}]{bowie2005solving}%
  \BibitemOpen
  \bibfield  {author} {\bibinfo {author} {\bibfnamefont {J.~U.}\ \bibnamefont
  {Bowie}},\ }\href@noop {} {\bibfield  {journal} {\bibinfo  {journal}
  {Nature}\ }\textbf {\bibinfo {volume} {438}},\ \bibinfo {pages} {581}
  (\bibinfo {year} {2005})}\BibitemShut {NoStop}%
\bibitem [{\citenamefont {Popot}\ and\ \citenamefont
  {Engelman}(1990)}]{popot1990membrane}%
  \BibitemOpen
  \bibfield  {author} {\bibinfo {author} {\bibfnamefont {J.-L.}\ \bibnamefont
  {Popot}}\ and\ \bibinfo {author} {\bibfnamefont {D.~M.}\ \bibnamefont
  {Engelman}},\ }\href@noop {} {\bibfield  {journal} {\bibinfo  {journal}
  {Biochemistry}\ }\textbf {\bibinfo {volume} {29}},\ \bibinfo {pages} {4031}
  (\bibinfo {year} {1990})}\BibitemShut {NoStop}%
\bibitem [{\citenamefont {Bowie}(2011)}]{bowie2011membrane}%
  \BibitemOpen
  \bibfield  {author} {\bibinfo {author} {\bibfnamefont {J.~U.}\ \bibnamefont
  {Bowie}},\ }\href@noop {} {\bibfield  {journal} {\bibinfo  {journal} {Current
  opinion in structural biology}\ }\textbf {\bibinfo {volume} {21}},\ \bibinfo
  {pages} {42} (\bibinfo {year} {2011})}\BibitemShut {NoStop}%
\bibitem [{\citenamefont {Killian}\ \emph {et~al.}(1996)\citenamefont
  {Killian}, \citenamefont {Salemink}, \citenamefont {de~Planque},
  \citenamefont {Lindblom}, \citenamefont {Koeppe},\ and\ \citenamefont
  {Greathouse}}]{killian1996induction}%
  \BibitemOpen
  \bibfield  {author} {\bibinfo {author} {\bibfnamefont {J.~A.}\ \bibnamefont
  {Killian}}, \bibinfo {author} {\bibfnamefont {I.}~\bibnamefont {Salemink}},
  \bibinfo {author} {\bibfnamefont {M.~R.}\ \bibnamefont {de~Planque}},
  \bibinfo {author} {\bibfnamefont {G.}~\bibnamefont {Lindblom}}, \bibinfo
  {author} {\bibfnamefont {R.~E.}\ \bibnamefont {Koeppe}}, \ and\ \bibinfo
  {author} {\bibfnamefont {D.~V.}\ \bibnamefont {Greathouse}},\ }\href@noop {}
  {\bibfield  {journal} {\bibinfo  {journal} {Biochemistry}\ }\textbf {\bibinfo
  {volume} {35}},\ \bibinfo {pages} {1037} (\bibinfo {year}
  {1996})}\BibitemShut {NoStop}%
\bibitem [{\citenamefont {Morein}\ \emph {et~al.}(1997)\citenamefont {Morein},
  \citenamefont {Strandberg}, \citenamefont {Killian}, \citenamefont {Persson},
  \citenamefont {Arvidson}, \citenamefont {Koeppe~2nd},\ and\ \citenamefont
  {Lindblom}}]{morein1997influence}%
  \BibitemOpen
  \bibfield  {author} {\bibinfo {author} {\bibfnamefont {S.}~\bibnamefont
  {Morein}}, \bibinfo {author} {\bibfnamefont {E.}~\bibnamefont {Strandberg}},
  \bibinfo {author} {\bibfnamefont {J.~A.}\ \bibnamefont {Killian}}, \bibinfo
  {author} {\bibfnamefont {S.}~\bibnamefont {Persson}}, \bibinfo {author}
  {\bibfnamefont {G.}~\bibnamefont {Arvidson}}, \bibinfo {author}
  {\bibfnamefont {R.}~\bibnamefont {Koeppe~2nd}}, \ and\ \bibinfo {author}
  {\bibfnamefont {G.}~\bibnamefont {Lindblom}},\ }\href@noop {} {\bibfield
  {journal} {\bibinfo  {journal} {Biophysical journal}\ }\textbf {\bibinfo
  {volume} {73}},\ \bibinfo {pages} {3078} (\bibinfo {year}
  {1997})}\BibitemShut {NoStop}%
\bibitem [{\citenamefont {Strandberg}\ \emph {et~al.}(2004)\citenamefont
  {Strandberg}, \citenamefont {{\"O}zdirekcan}, \citenamefont {Rijkers},
  \citenamefont {van~der Wel}, \citenamefont {Koeppe}, \citenamefont
  {Liskamp},\ and\ \citenamefont {Killian}}]{strandberg2004tilt}%
  \BibitemOpen
  \bibfield  {author} {\bibinfo {author} {\bibfnamefont {E.}~\bibnamefont
  {Strandberg}}, \bibinfo {author} {\bibfnamefont {S.}~\bibnamefont
  {{\"O}zdirekcan}}, \bibinfo {author} {\bibfnamefont {D.~T.}\ \bibnamefont
  {Rijkers}}, \bibinfo {author} {\bibfnamefont {P.~C.}\ \bibnamefont {van~der
  Wel}}, \bibinfo {author} {\bibfnamefont {R.~E.}\ \bibnamefont {Koeppe}},
  \bibinfo {author} {\bibfnamefont {R.~M.}\ \bibnamefont {Liskamp}}, \ and\
  \bibinfo {author} {\bibfnamefont {J.~A.}\ \bibnamefont {Killian}},\
  }\href@noop {} {\bibfield  {journal} {\bibinfo  {journal} {Biophysical
  journal}\ }\textbf {\bibinfo {volume} {86}},\ \bibinfo {pages} {3709}
  (\bibinfo {year} {2004})}\BibitemShut {NoStop}%
\bibitem [{\citenamefont {Im}\ and\ \citenamefont
  {Brooks}(2005)}]{im2005interfacial}%
  \BibitemOpen
  \bibfield  {author} {\bibinfo {author} {\bibfnamefont {W.}~\bibnamefont
  {Im}}\ and\ \bibinfo {author} {\bibfnamefont {C.~L.}\ \bibnamefont
  {Brooks}},\ }\href@noop {} {\bibfield  {journal} {\bibinfo  {journal}
  {Proceedings of the National Academy of Sciences of the United States of
  America}\ }\textbf {\bibinfo {volume} {102}},\ \bibinfo {pages} {6771}
  (\bibinfo {year} {2005})}\BibitemShut {NoStop}%
\bibitem [{\citenamefont {Kandasamy}\ and\ \citenamefont
  {Larson}(2006)}]{kandasamy2006molecular}%
  \BibitemOpen
  \bibfield  {author} {\bibinfo {author} {\bibfnamefont {S.~K.}\ \bibnamefont
  {Kandasamy}}\ and\ \bibinfo {author} {\bibfnamefont {R.~G.}\ \bibnamefont
  {Larson}},\ }\href@noop {} {\bibfield  {journal} {\bibinfo  {journal}
  {Biophysical journal}\ }\textbf {\bibinfo {volume} {90}},\ \bibinfo {pages}
  {2326} (\bibinfo {year} {2006})}\BibitemShut {NoStop}%
\bibitem [{\citenamefont {{\"O}zdirekcan}\ \emph {et~al.}(2007)\citenamefont
  {{\"O}zdirekcan}, \citenamefont {Etchebest}, \citenamefont {Killian},\ and\
  \citenamefont {Fuchs}}]{ozdirekcan2007orientation}%
  \BibitemOpen
  \bibfield  {author} {\bibinfo {author} {\bibfnamefont {S.}~\bibnamefont
  {{\"O}zdirekcan}}, \bibinfo {author} {\bibfnamefont {C.}~\bibnamefont
  {Etchebest}}, \bibinfo {author} {\bibfnamefont {J.~A.}\ \bibnamefont
  {Killian}}, \ and\ \bibinfo {author} {\bibfnamefont {P.~F.}\ \bibnamefont
  {Fuchs}},\ }\href@noop {} {\bibfield  {journal} {\bibinfo  {journal} {Journal
  of the American Chemical Society}\ }\textbf {\bibinfo {volume} {129}},\
  \bibinfo {pages} {15174} (\bibinfo {year} {2007})}\BibitemShut {NoStop}%
\bibitem [{\citenamefont {Monticelli}, \citenamefont {Tieleman},\ and\
  \citenamefont {Fuchs}(2010)}]{monticelli2010interpretation}%
  \BibitemOpen
  \bibfield  {author} {\bibinfo {author} {\bibfnamefont {L.}~\bibnamefont
  {Monticelli}}, \bibinfo {author} {\bibfnamefont {D.~P.}\ \bibnamefont
  {Tieleman}}, \ and\ \bibinfo {author} {\bibfnamefont {P.~F.}\ \bibnamefont
  {Fuchs}},\ }\href@noop {} {\bibfield  {journal} {\bibinfo  {journal}
  {Biophysical journal}\ }\textbf {\bibinfo {volume} {99}},\ \bibinfo {pages}
  {1455} (\bibinfo {year} {2010})}\BibitemShut {NoStop}%
\bibitem [{\citenamefont {Kim}\ and\ \citenamefont
  {Im}(2010)}]{kim2010revisiting}%
  \BibitemOpen
  \bibfield  {author} {\bibinfo {author} {\bibfnamefont {T.}~\bibnamefont
  {Kim}}\ and\ \bibinfo {author} {\bibfnamefont {W.}~\bibnamefont {Im}},\
  }\href@noop {} {\bibfield  {journal} {\bibinfo  {journal} {Biophysical
  journal}\ }\textbf {\bibinfo {volume} {99}},\ \bibinfo {pages} {175}
  (\bibinfo {year} {2010})}\BibitemShut {NoStop}%
\bibitem [{\citenamefont {Nymeyer}, \citenamefont {Woolf},\ and\ \citenamefont
  {Garcia}(2005)}]{nymeyer2005folding}%
  \BibitemOpen
  \bibfield  {author} {\bibinfo {author} {\bibfnamefont {H.}~\bibnamefont
  {Nymeyer}}, \bibinfo {author} {\bibfnamefont {T.~B.}\ \bibnamefont {Woolf}},
  \ and\ \bibinfo {author} {\bibfnamefont {A.~E.}\ \bibnamefont {Garcia}},\
  }\href@noop {} {\bibfield  {journal} {\bibinfo  {journal} {Proteins: Struct.,
  Funct., Bioinf.}\ }\textbf {\bibinfo {volume} {59}},\ \bibinfo {pages} {783}
  (\bibinfo {year} {2005})}\BibitemShut {NoStop}%
\bibitem [{\citenamefont {Ulmschneider}\ \emph {et~al.}(2010)\citenamefont
  {Ulmschneider}, \citenamefont {Doux}, \citenamefont {Killian}, \citenamefont
  {Smith},\ and\ \citenamefont {Ulmschneider}}]{ulmschneider2010mechanism}%
  \BibitemOpen
  \bibfield  {author} {\bibinfo {author} {\bibfnamefont {M.~B.}\ \bibnamefont
  {Ulmschneider}}, \bibinfo {author} {\bibfnamefont {J.~P.}\ \bibnamefont
  {Doux}}, \bibinfo {author} {\bibfnamefont {J.~A.}\ \bibnamefont {Killian}},
  \bibinfo {author} {\bibfnamefont {J.~C.}\ \bibnamefont {Smith}}, \ and\
  \bibinfo {author} {\bibfnamefont {J.~P.}\ \bibnamefont {Ulmschneider}},\
  }\href@noop {} {\bibfield  {journal} {\bibinfo  {journal} {Journal of the
  American Chemical Society}\ }\textbf {\bibinfo {volume} {132}},\ \bibinfo
  {pages} {3452} (\bibinfo {year} {2010})}\BibitemShut {NoStop}%
\bibitem [{\citenamefont {Bereau}, \citenamefont {Wang},\ and\ \citenamefont
  {Deserno}(2014)}]{bereau2014more}%
  \BibitemOpen
  \bibfield  {author} {\bibinfo {author} {\bibfnamefont {T.}~\bibnamefont
  {Bereau}}, \bibinfo {author} {\bibfnamefont {Z.-J.}\ \bibnamefont {Wang}}, \
  and\ \bibinfo {author} {\bibfnamefont {M.}~\bibnamefont {Deserno}},\
  }\href@noop {} {\bibfield  {journal} {\bibinfo  {journal} {J. Chem. Phys.}\
  }\textbf {\bibinfo {volume} {140}},\ \bibinfo {pages} {115101} (\bibinfo
  {year} {2014})}\BibitemShut {NoStop}%
\bibitem [{\citenamefont {Bereau}\ and\ \citenamefont
  {Deserno}(2014)}]{bereau2014enhanced}%
  \BibitemOpen
  \bibfield  {author} {\bibinfo {author} {\bibfnamefont {T.}~\bibnamefont
  {Bereau}}\ and\ \bibinfo {author} {\bibfnamefont {M.}~\bibnamefont
  {Deserno}},\ }\href@noop {} {\bibfield  {journal} {\bibinfo  {journal} {The
  Journal of membrane biology}\ }\textbf {\bibinfo {volume} {248}},\ \bibinfo
  {pages} {395} (\bibinfo {year} {2014})}\BibitemShut {NoStop}%
\bibitem [{\citenamefont {Marrink}\ \emph {et~al.}(2007)\citenamefont
  {Marrink}, \citenamefont {Risselada}, \citenamefont {Yefimov}, \citenamefont
  {Tieleman},\ and\ \citenamefont {De~Vries}}]{marrink2007martini}%
  \BibitemOpen
  \bibfield  {author} {\bibinfo {author} {\bibfnamefont {S.~J.}\ \bibnamefont
  {Marrink}}, \bibinfo {author} {\bibfnamefont {H.~J.}\ \bibnamefont
  {Risselada}}, \bibinfo {author} {\bibfnamefont {S.}~\bibnamefont {Yefimov}},
  \bibinfo {author} {\bibfnamefont {D.~P.}\ \bibnamefont {Tieleman}}, \ and\
  \bibinfo {author} {\bibfnamefont {A.~H.}\ \bibnamefont {De~Vries}},\
  }\href@noop {} {\bibfield  {journal} {\bibinfo  {journal} {The Journal of
  Physical Chemistry B}\ }\textbf {\bibinfo {volume} {111}},\ \bibinfo {pages}
  {7812} (\bibinfo {year} {2007})}\BibitemShut {NoStop}%
\bibitem [{\citenamefont {Bond}, \citenamefont {Wee},\ and\ \citenamefont
  {Sansom}(2008)}]{bond2008coarse}%
  \BibitemOpen
  \bibfield  {author} {\bibinfo {author} {\bibfnamefont {P.~J.}\ \bibnamefont
  {Bond}}, \bibinfo {author} {\bibfnamefont {C.~L.}\ \bibnamefont {Wee}}, \
  and\ \bibinfo {author} {\bibfnamefont {M.~S.}\ \bibnamefont {Sansom}},\
  }\href@noop {} {\bibfield  {journal} {\bibinfo  {journal} {Biochemistry}\
  }\textbf {\bibinfo {volume} {47}},\ \bibinfo {pages} {11321} (\bibinfo {year}
  {2008})}\BibitemShut {NoStop}%
\bibitem [{\citenamefont {Bereau}\ and\ \citenamefont
  {Deserno}(2009)}]{bereau2009generic}%
  \BibitemOpen
  \bibfield  {author} {\bibinfo {author} {\bibfnamefont {T.}~\bibnamefont
  {Bereau}}\ and\ \bibinfo {author} {\bibfnamefont {M.}~\bibnamefont
  {Deserno}},\ }\href@noop {} {\bibfield  {journal} {\bibinfo  {journal} {J.
  Chem. Phys.}\ }\textbf {\bibinfo {volume} {130}},\ \bibinfo {pages} {235106}
  (\bibinfo {year} {2009})}\BibitemShut {NoStop}%
\bibitem [{\citenamefont {Wang}\ and\ \citenamefont
  {Deserno}(2010{\natexlab{a}})}]{wang2010systematically}%
  \BibitemOpen
  \bibfield  {author} {\bibinfo {author} {\bibfnamefont {Z.-J.}\ \bibnamefont
  {Wang}}\ and\ \bibinfo {author} {\bibfnamefont {M.}~\bibnamefont {Deserno}},\
  }\href@noop {} {\bibfield  {journal} {\bibinfo  {journal} {J. Phys. Chem. B}\
  }\textbf {\bibinfo {volume} {114}},\ \bibinfo {pages} {11207} (\bibinfo
  {year} {2010}{\natexlab{a}})}\BibitemShut {NoStop}%
\bibitem [{\citenamefont {Wang}\ and\ \citenamefont
  {Deserno}(2010{\natexlab{b}})}]{wang2010systematic}%
  \BibitemOpen
  \bibfield  {author} {\bibinfo {author} {\bibfnamefont {Z.-J.}\ \bibnamefont
  {Wang}}\ and\ \bibinfo {author} {\bibfnamefont {M.}~\bibnamefont {Deserno}},\
  }\href@noop {} {\bibfield  {journal} {\bibinfo  {journal} {New J. Physics}\
  }\textbf {\bibinfo {volume} {12}},\ \bibinfo {pages} {095004} (\bibinfo
  {year} {2010}{\natexlab{b}})}\BibitemShut {NoStop}%
\bibitem [{\citenamefont {Bereau}, \citenamefont {Bachmann},\ and\
  \citenamefont {Deserno}(2010)}]{bereau2010interplay}%
  \BibitemOpen
  \bibfield  {author} {\bibinfo {author} {\bibfnamefont {T.}~\bibnamefont
  {Bereau}}, \bibinfo {author} {\bibfnamefont {M.}~\bibnamefont {Bachmann}}, \
  and\ \bibinfo {author} {\bibfnamefont {M.}~\bibnamefont {Deserno}},\
  }\href@noop {} {\bibfield  {journal} {\bibinfo  {journal} {J. Am. Chem.
  Soc.}\ }\textbf {\bibinfo {volume} {132}},\ \bibinfo {pages} {13129}
  (\bibinfo {year} {2010})}\BibitemShut {NoStop}%
\bibitem [{\citenamefont {Bereau}, \citenamefont {Deserno},\ and\ \citenamefont
  {Bachmann}(2011)}]{bereau2011structural}%
  \BibitemOpen
  \bibfield  {author} {\bibinfo {author} {\bibfnamefont {T.}~\bibnamefont
  {Bereau}}, \bibinfo {author} {\bibfnamefont {M.}~\bibnamefont {Deserno}}, \
  and\ \bibinfo {author} {\bibfnamefont {M.}~\bibnamefont {Bachmann}},\
  }\href@noop {} {\bibfield  {journal} {\bibinfo  {journal} {Biophys. J.}\
  }\textbf {\bibinfo {volume} {100}},\ \bibinfo {pages} {2764} (\bibinfo {year}
  {2011})}\BibitemShut {NoStop}%
\bibitem [{\citenamefont {Osborne}, \citenamefont {Bachmann},\ and\
  \citenamefont {Strodel}(2013)}]{osborne2013thermodynamic}%
  \BibitemOpen
  \bibfield  {author} {\bibinfo {author} {\bibfnamefont {K.~L.}\ \bibnamefont
  {Osborne}}, \bibinfo {author} {\bibfnamefont {M.}~\bibnamefont {Bachmann}}, \
  and\ \bibinfo {author} {\bibfnamefont {B.}~\bibnamefont {Strodel}},\
  }\href@noop {} {\bibfield  {journal} {\bibinfo  {journal} {Proteins:
  Structure, Function, and Bioinformatics}\ }\textbf {\bibinfo {volume} {81}},\
  \bibinfo {pages} {1141} (\bibinfo {year} {2013})}\BibitemShut {NoStop}%
\bibitem [{\citenamefont {Bereau}\ \emph {et~al.}(2012)\citenamefont {Bereau},
  \citenamefont {Globisch}, \citenamefont {Deserno},\ and\ \citenamefont
  {Peter}}]{bereau2012coarse}%
  \BibitemOpen
  \bibfield  {author} {\bibinfo {author} {\bibfnamefont {T.}~\bibnamefont
  {Bereau}}, \bibinfo {author} {\bibfnamefont {C.}~\bibnamefont {Globisch}},
  \bibinfo {author} {\bibfnamefont {M.}~\bibnamefont {Deserno}}, \ and\
  \bibinfo {author} {\bibfnamefont {C.}~\bibnamefont {Peter}},\ }\href@noop {}
  {\bibfield  {journal} {\bibinfo  {journal} {J. Chem. Theory Comput.}\
  }\textbf {\bibinfo {volume} {8}},\ \bibinfo {pages} {3750} (\bibinfo {year}
  {2012})}\BibitemShut {NoStop}%
\bibitem [{\citenamefont {Reith}, \citenamefont {P{\"u}tz},\ and\ \citenamefont
  {M{\"u}ller-Plathe}(2003)}]{reith2003deriving}%
  \BibitemOpen
  \bibfield  {author} {\bibinfo {author} {\bibfnamefont {D.}~\bibnamefont
  {Reith}}, \bibinfo {author} {\bibfnamefont {M.}~\bibnamefont {P{\"u}tz}}, \
  and\ \bibinfo {author} {\bibfnamefont {F.}~\bibnamefont
  {M{\"u}ller-Plathe}},\ }\href@noop {} {\bibfield  {journal} {\bibinfo
  {journal} {Journal of computational chemistry}\ }\textbf {\bibinfo {volume}
  {24}},\ \bibinfo {pages} {1624} (\bibinfo {year} {2003})}\BibitemShut
  {NoStop}%
\bibitem [{\citenamefont {MacCallum}, \citenamefont {Bennett},\ and\
  \citenamefont {Tieleman}(2008)}]{maccallum2008distribution}%
  \BibitemOpen
  \bibfield  {author} {\bibinfo {author} {\bibfnamefont {J.~L.}\ \bibnamefont
  {MacCallum}}, \bibinfo {author} {\bibfnamefont {W.}~\bibnamefont {Bennett}},
  \ and\ \bibinfo {author} {\bibfnamefont {D.~P.}\ \bibnamefont {Tieleman}},\
  }\href@noop {} {\bibfield  {journal} {\bibinfo  {journal} {Biophys. J.}\
  }\textbf {\bibinfo {volume} {94}},\ \bibinfo {pages} {3393} (\bibinfo {year}
  {2008})}\BibitemShut {NoStop}%
\bibitem [{\citenamefont {de~Jong}\ \emph {et~al.}(2012)\citenamefont
  {de~Jong}, \citenamefont {Singh}, \citenamefont {Bennett}, \citenamefont
  {Arnarez}, \citenamefont {Wassenaar}, \citenamefont {Schafer}, \citenamefont
  {Periole}, \citenamefont {Tieleman},\ and\ \citenamefont
  {Marrink}}]{de2012improved}%
  \BibitemOpen
  \bibfield  {author} {\bibinfo {author} {\bibfnamefont {D.~H.}\ \bibnamefont
  {de~Jong}}, \bibinfo {author} {\bibfnamefont {G.}~\bibnamefont {Singh}},
  \bibinfo {author} {\bibfnamefont {W.~D.}\ \bibnamefont {Bennett}}, \bibinfo
  {author} {\bibfnamefont {C.}~\bibnamefont {Arnarez}}, \bibinfo {author}
  {\bibfnamefont {T.~A.}\ \bibnamefont {Wassenaar}}, \bibinfo {author}
  {\bibfnamefont {L.~V.}\ \bibnamefont {Schafer}}, \bibinfo {author}
  {\bibfnamefont {X.}~\bibnamefont {Periole}}, \bibinfo {author} {\bibfnamefont
  {D.~P.}\ \bibnamefont {Tieleman}}, \ and\ \bibinfo {author} {\bibfnamefont
  {S.~J.}\ \bibnamefont {Marrink}},\ }\href@noop {} {\bibfield  {journal}
  {\bibinfo  {journal} {Journal of Chemical Theory and Computation}\ }\textbf
  {\bibinfo {volume} {9}},\ \bibinfo {pages} {687} (\bibinfo {year}
  {2012})}\BibitemShut {NoStop}%
\bibitem [{\citenamefont {Monticelli}\ \emph {et~al.}(2008)\citenamefont
  {Monticelli}, \citenamefont {Kandasamy}, \citenamefont {Periole},
  \citenamefont {Larson}, \citenamefont {Tieleman},\ and\ \citenamefont
  {Marrink}}]{monticelli2008martini}%
  \BibitemOpen
  \bibfield  {author} {\bibinfo {author} {\bibfnamefont {L.}~\bibnamefont
  {Monticelli}}, \bibinfo {author} {\bibfnamefont {S.~K.}\ \bibnamefont
  {Kandasamy}}, \bibinfo {author} {\bibfnamefont {X.}~\bibnamefont {Periole}},
  \bibinfo {author} {\bibfnamefont {R.~G.}\ \bibnamefont {Larson}}, \bibinfo
  {author} {\bibfnamefont {D.~P.}\ \bibnamefont {Tieleman}}, \ and\ \bibinfo
  {author} {\bibfnamefont {S.-J.}\ \bibnamefont {Marrink}},\ }\href@noop {}
  {\bibfield  {journal} {\bibinfo  {journal} {Journal of chemical theory and
  computation}\ }\textbf {\bibinfo {volume} {4}},\ \bibinfo {pages} {819}
  (\bibinfo {year} {2008})}\BibitemShut {NoStop}%
\bibitem [{\citenamefont {L{\'o}pez}\ \emph {et~al.}(2009)\citenamefont
  {L{\'o}pez}, \citenamefont {Rzepiela}, \citenamefont {De~Vries},
  \citenamefont {Dijkhuizen}, \citenamefont {Hünenberger},\ and\
  \citenamefont {Marrink}}]{lopez2009martini}%
  \BibitemOpen
  \bibfield  {author} {\bibinfo {author} {\bibfnamefont {C.~A.}\ \bibnamefont
  {L{\'o}pez}}, \bibinfo {author} {\bibfnamefont {A.~J.}\ \bibnamefont
  {Rzepiela}}, \bibinfo {author} {\bibfnamefont {A.~H.}\ \bibnamefont
  {De~Vries}}, \bibinfo {author} {\bibfnamefont {L.}~\bibnamefont
  {Dijkhuizen}}, \bibinfo {author} {\bibfnamefont {P.~H.}\ \bibnamefont
  {Hünenberger}}, \ and\ \bibinfo {author} {\bibfnamefont {S.~J.}\
  \bibnamefont {Marrink}},\ }\href@noop {} {\bibfield  {journal} {\bibinfo
  {journal} {Journal of Chemical Theory and Computation}\ }\textbf {\bibinfo
  {volume} {5}},\ \bibinfo {pages} {3195} (\bibinfo {year} {2009})}\BibitemShut
  {NoStop}%
\bibitem [{\citenamefont {Bereau}\ and\ \citenamefont
  {Kremer}(2015)}]{bereau2015automated}%
  \BibitemOpen
  \bibfield  {author} {\bibinfo {author} {\bibfnamefont {T.}~\bibnamefont
  {Bereau}}\ and\ \bibinfo {author} {\bibfnamefont {K.}~\bibnamefont
  {Kremer}},\ }\href@noop {} {\bibfield  {journal} {\bibinfo  {journal}
  {Journal of Chemical Theory and Computation}\ } (\bibinfo {year}
  {2015})}\BibitemShut {NoStop}%
\bibitem [{\citenamefont {Ing{\'o}lfsson}\ \emph {et~al.}(2014)\citenamefont
  {Ing{\'o}lfsson}, \citenamefont {Melo}, \citenamefont {van Eerden},
  \citenamefont {Arnarez}, \citenamefont {Lopez}, \citenamefont {Wassenaar},
  \citenamefont {Periole}, \citenamefont {De~Vries}, \citenamefont {Tieleman},\
  and\ \citenamefont {Marrink}}]{ingolfsson2014lipid}%
  \BibitemOpen
  \bibfield  {author} {\bibinfo {author} {\bibfnamefont {H.~I.}\ \bibnamefont
  {Ing{\'o}lfsson}}, \bibinfo {author} {\bibfnamefont {M.~N.}\ \bibnamefont
  {Melo}}, \bibinfo {author} {\bibfnamefont {F.~J.}\ \bibnamefont {van
  Eerden}}, \bibinfo {author} {\bibfnamefont {C.}~\bibnamefont {Arnarez}},
  \bibinfo {author} {\bibfnamefont {C.~A.}\ \bibnamefont {Lopez}}, \bibinfo
  {author} {\bibfnamefont {T.~A.}\ \bibnamefont {Wassenaar}}, \bibinfo {author}
  {\bibfnamefont {X.}~\bibnamefont {Periole}}, \bibinfo {author} {\bibfnamefont
  {A.~H.}\ \bibnamefont {De~Vries}}, \bibinfo {author} {\bibfnamefont {D.~P.}\
  \bibnamefont {Tieleman}}, \ and\ \bibinfo {author} {\bibfnamefont {S.~J.}\
  \bibnamefont {Marrink}},\ }\href@noop {} {\bibfield  {journal} {\bibinfo
  {journal} {Journal of the American Chemical Society}\ }\textbf {\bibinfo
  {volume} {136}},\ \bibinfo {pages} {14554} (\bibinfo {year}
  {2014})}\BibitemShut {NoStop}%
\bibitem [{\citenamefont {Wassenaar}\ \emph {et~al.}(2015)\citenamefont
  {Wassenaar}, \citenamefont {Ing{\'o}lfsson}, \citenamefont {Böckmann},
  \citenamefont {Tieleman},\ and\ \citenamefont
  {Marrink}}]{wassenaar2015computational}%
  \BibitemOpen
  \bibfield  {author} {\bibinfo {author} {\bibfnamefont {T.~A.}\ \bibnamefont
  {Wassenaar}}, \bibinfo {author} {\bibfnamefont {H.~I.}\ \bibnamefont
  {Ing{\'o}lfsson}}, \bibinfo {author} {\bibfnamefont {R.~A.}\ \bibnamefont
  {Böckmann}}, \bibinfo {author} {\bibfnamefont {D.~P.}\ \bibnamefont
  {Tieleman}}, \ and\ \bibinfo {author} {\bibfnamefont {S.~J.}\ \bibnamefont
  {Marrink}},\ }\href@noop {} {\bibfield  {journal} {\bibinfo  {journal}
  {Journal of Chemical Theory and Computation}\ }\textbf {\bibinfo {volume}
  {11}},\ \bibinfo {pages} {2144} (\bibinfo {year} {2015})}\BibitemShut
  {NoStop}%
\bibitem [{\citenamefont {Kopelevich}(2013)}]{kopelevich2013one}%
  \BibitemOpen
  \bibfield  {author} {\bibinfo {author} {\bibfnamefont {D.~I.}\ \bibnamefont
  {Kopelevich}},\ }\href@noop {} {\bibfield  {journal} {\bibinfo  {journal}
  {The Journal of chemical physics}\ }\textbf {\bibinfo {volume} {139}},\
  \bibinfo {pages} {134906} (\bibinfo {year} {2013})}\BibitemShut {NoStop}%
\bibitem [{\citenamefont {Humphrey}, \citenamefont {Dalke},\ and\ \citenamefont
  {Schulten}(1996)}]{humphrey1996vmd}%
  \BibitemOpen
  \bibfield  {author} {\bibinfo {author} {\bibfnamefont {W.}~\bibnamefont
  {Humphrey}}, \bibinfo {author} {\bibfnamefont {A.}~\bibnamefont {Dalke}}, \
  and\ \bibinfo {author} {\bibfnamefont {K.}~\bibnamefont {Schulten}},\
  }\href@noop {} {\bibfield  {journal} {\bibinfo  {journal} {J. Mol. Graphics}\
  }\textbf {\bibinfo {volume} {14}},\ \bibinfo {pages} {33} (\bibinfo {year}
  {1996})}\BibitemShut {NoStop}%
\bibitem [{\citenamefont {Ramadurai}\ \emph {et~al.}(2010)\citenamefont
  {Ramadurai}, \citenamefont {Holt}, \citenamefont {Sch{\"a}fer}, \citenamefont
  {Krasnikov}, \citenamefont {Rijkers}, \citenamefont {Marrink}, \citenamefont
  {Killian},\ and\ \citenamefont {Poolman}}]{ramadurai2010influence}%
  \BibitemOpen
  \bibfield  {author} {\bibinfo {author} {\bibfnamefont {S.}~\bibnamefont
  {Ramadurai}}, \bibinfo {author} {\bibfnamefont {A.}~\bibnamefont {Holt}},
  \bibinfo {author} {\bibfnamefont {L.~V.}\ \bibnamefont {Sch{\"a}fer}},
  \bibinfo {author} {\bibfnamefont {V.~V.}\ \bibnamefont {Krasnikov}}, \bibinfo
  {author} {\bibfnamefont {D.~T.}\ \bibnamefont {Rijkers}}, \bibinfo {author}
  {\bibfnamefont {S.~J.}\ \bibnamefont {Marrink}}, \bibinfo {author}
  {\bibfnamefont {J.~A.}\ \bibnamefont {Killian}}, \ and\ \bibinfo {author}
  {\bibfnamefont {B.}~\bibnamefont {Poolman}},\ }\href@noop {} {\bibfield
  {journal} {\bibinfo  {journal} {Biophysical journal}\ }\textbf {\bibinfo
  {volume} {99}},\ \bibinfo {pages} {1447} (\bibinfo {year}
  {2010})}\BibitemShut {NoStop}%
\bibitem [{\citenamefont {Jorgensen}, \citenamefont {Maxwell},\ and\
  \citenamefont {Tirado-Rives}(1996)}]{jorgensen1996development}%
  \BibitemOpen
  \bibfield  {author} {\bibinfo {author} {\bibfnamefont {W.~L.}\ \bibnamefont
  {Jorgensen}}, \bibinfo {author} {\bibfnamefont {D.~S.}\ \bibnamefont
  {Maxwell}}, \ and\ \bibinfo {author} {\bibfnamefont {J.}~\bibnamefont
  {Tirado-Rives}},\ }\href@noop {} {\bibfield  {journal} {\bibinfo  {journal}
  {Journal of the American Chemical Society}\ }\textbf {\bibinfo {volume}
  {118}},\ \bibinfo {pages} {11225} (\bibinfo {year} {1996})}\BibitemShut
  {NoStop}%
\bibitem [{\citenamefont {Hall}, \citenamefont {Chetwynd},\ and\ \citenamefont
  {Sansom}(2011)}]{hall2011exploring}%
  \BibitemOpen
  \bibfield  {author} {\bibinfo {author} {\bibfnamefont {B.~A.}\ \bibnamefont
  {Hall}}, \bibinfo {author} {\bibfnamefont {A.~P.}\ \bibnamefont {Chetwynd}},
  \ and\ \bibinfo {author} {\bibfnamefont {M.~S.}\ \bibnamefont {Sansom}},\
  }\href@noop {} {\bibfield  {journal} {\bibinfo  {journal} {Biophysical
  journal}\ }\textbf {\bibinfo {volume} {100}},\ \bibinfo {pages} {1940}
  (\bibinfo {year} {2011})}\BibitemShut {NoStop}%
\bibitem [{\citenamefont {Neale}\ \emph {et~al.}(2011)\citenamefont {Neale},
  \citenamefont {Bennett}, \citenamefont {Tieleman},\ and\ \citenamefont
  {Pom{\`e}s}}]{neale2011statistical}%
  \BibitemOpen
  \bibfield  {author} {\bibinfo {author} {\bibfnamefont {C.}~\bibnamefont
  {Neale}}, \bibinfo {author} {\bibfnamefont {W.~D.}\ \bibnamefont {Bennett}},
  \bibinfo {author} {\bibfnamefont {D.~P.}\ \bibnamefont {Tieleman}}, \ and\
  \bibinfo {author} {\bibfnamefont {R.}~\bibnamefont {Pom{\`e}s}},\ }\href@noop
  {} {\bibfield  {journal} {\bibinfo  {journal} {J. Chem. Theory Comput.}\
  }\textbf {\bibinfo {volume} {7}},\ \bibinfo {pages} {4175} (\bibinfo {year}
  {2011})}\BibitemShut {NoStop}%
\bibitem [{\citenamefont {Laio}\ and\ \citenamefont
  {Parrinello}(2002)}]{laio2002escaping}%
  \BibitemOpen
  \bibfield  {author} {\bibinfo {author} {\bibfnamefont {A.}~\bibnamefont
  {Laio}}\ and\ \bibinfo {author} {\bibfnamefont {M.}~\bibnamefont
  {Parrinello}},\ }\href@noop {} {\bibfield  {journal} {\bibinfo  {journal}
  {Proceedings of the National Academy of Sciences}\ }\textbf {\bibinfo
  {volume} {99}},\ \bibinfo {pages} {12562} (\bibinfo {year}
  {2002})}\BibitemShut {NoStop}%
\bibitem [{\citenamefont {Limbach}\ \emph {et~al.}(2006)\citenamefont
  {Limbach}, \citenamefont {Arnold}, \citenamefont {Mann},\ and\ \citenamefont
  {Holm}}]{espresso}%
  \BibitemOpen
  \bibfield  {author} {\bibinfo {author} {\bibfnamefont {H.-J.}\ \bibnamefont
  {Limbach}}, \bibinfo {author} {\bibfnamefont {A.}~\bibnamefont {Arnold}},
  \bibinfo {author} {\bibfnamefont {B.~A.}\ \bibnamefont {Mann}}, \ and\
  \bibinfo {author} {\bibfnamefont {C.}~\bibnamefont {Holm}},\ }\href@noop {}
  {\bibfield  {journal} {\bibinfo  {journal} {Comput. Phys. Comm.}\ }\textbf
  {\bibinfo {volume} {174}},\ \bibinfo {pages} {704} (\bibinfo {year}
  {2006})}\BibitemShut {NoStop}%
\bibitem [{\citenamefont {Kolb}\ and\ \citenamefont
  {D{\"u}nweg}(1999)}]{kolb1999optimized}%
  \BibitemOpen
  \bibfield  {author} {\bibinfo {author} {\bibfnamefont {A.}~\bibnamefont
  {Kolb}}\ and\ \bibinfo {author} {\bibfnamefont {B.}~\bibnamefont
  {D{\"u}nweg}},\ }\href@noop {} {\bibfield  {journal} {\bibinfo  {journal} {J.
  Chem. Phys.}\ }\textbf {\bibinfo {volume} {111}},\ \bibinfo {pages} {4453}
  (\bibinfo {year} {1999})}\BibitemShut {NoStop}%
\bibitem [{\citenamefont {Bereau}(2015)}]{bereau2015multi}%
  \BibitemOpen
  \bibfield  {author} {\bibinfo {author} {\bibfnamefont {T.}~\bibnamefont
  {Bereau}},\ }\href@noop {} {\bibfield  {journal} {\bibinfo  {journal}
  {Physics Procedia}\ }\textbf {\bibinfo {volume} {68}},\ \bibinfo {pages} {7}
  (\bibinfo {year} {2015})},\ \bibinfo {note} {{DOI:
  10.1016/j.phpro.2015.07.101}}\BibitemShut {NoStop}%
\bibitem [{\citenamefont {Frishman}\ and\ \citenamefont
  {Argos}(1995)}]{stride}%
  \BibitemOpen
  \bibfield  {author} {\bibinfo {author} {\bibfnamefont {D.}~\bibnamefont
  {Frishman}}\ and\ \bibinfo {author} {\bibfnamefont {P.}~\bibnamefont
  {Argos}},\ }\href@noop {} {\bibfield  {journal} {\bibinfo  {journal}
  {Proteins: Struct. Func. Genet.}\ }\textbf {\bibinfo {volume} {23}},\
  \bibinfo {pages} {566} (\bibinfo {year} {1995})}\BibitemShut {NoStop}%
\bibitem [{\citenamefont {Bunker}\ and\ \citenamefont
  {D{\"u}nweg}(2000)}]{bunker2000parallel}%
  \BibitemOpen
  \bibfield  {author} {\bibinfo {author} {\bibfnamefont {A.}~\bibnamefont
  {Bunker}}\ and\ \bibinfo {author} {\bibfnamefont {B.}~\bibnamefont
  {D{\"u}nweg}},\ }\href@noop {} {\bibfield  {journal} {\bibinfo  {journal}
  {Phys. Rev. E}\ }\textbf {\bibinfo {volume} {63}},\ \bibinfo {pages} {016701}
  (\bibinfo {year} {2000})}\BibitemShut {NoStop}%
\bibitem [{\citenamefont {Torrie}\ and\ \citenamefont
  {Valleau}(1977)}]{Torrie1977}%
  \BibitemOpen
  \bibfield  {author} {\bibinfo {author} {\bibfnamefont {G.~M.}\ \bibnamefont
  {Torrie}}\ and\ \bibinfo {author} {\bibfnamefont {J.~P.}\ \bibnamefont
  {Valleau}},\ }\href@noop {} {\bibfield  {journal} {\bibinfo  {journal} {J.
  Comput. Phys.}\ }\textbf {\bibinfo {volume} {23}},\ \bibinfo {pages} {187}
  (\bibinfo {year} {1977})}\BibitemShut {NoStop}%
\bibitem [{\citenamefont {Ferrenberg}\ and\ \citenamefont
  {Swendsen}(1989)}]{FeSw89}%
  \BibitemOpen
  \bibfield  {author} {\bibinfo {author} {\bibfnamefont {A.~M.}\ \bibnamefont
  {Ferrenberg}}\ and\ \bibinfo {author} {\bibfnamefont {R.~H.}\ \bibnamefont
  {Swendsen}},\ }\href {\doibase 10.1103/PhysRevLett.63.1195} {\bibfield
  {journal} {\bibinfo  {journal} {Phys. Rev. Lett.}\ }\textbf {\bibinfo
  {volume} {63}},\ \bibinfo {pages} {1195} (\bibinfo {year}
  {1989})}\BibitemShut {NoStop}%
\bibitem [{\citenamefont {Kumar}\ \emph {et~al.}(1992)\citenamefont {Kumar},
  \citenamefont {Rosenberg}, \citenamefont {Bouzida}, \citenamefont
  {Swendsen},\ and\ \citenamefont {Kollman}}]{Kumar92}%
  \BibitemOpen
  \bibfield  {author} {\bibinfo {author} {\bibfnamefont {S.}~\bibnamefont
  {Kumar}}, \bibinfo {author} {\bibfnamefont {J.~M.}\ \bibnamefont
  {Rosenberg}}, \bibinfo {author} {\bibfnamefont {D.}~\bibnamefont {Bouzida}},
  \bibinfo {author} {\bibfnamefont {R.~H.}\ \bibnamefont {Swendsen}}, \ and\
  \bibinfo {author} {\bibfnamefont {P.~A.}\ \bibnamefont {Kollman}},\ }\href
  {\doibase 10.1002/jcc.540130812} {\bibfield  {journal} {\bibinfo  {journal}
  {J. Comput. Chem.}\ }\textbf {\bibinfo {volume} {13}},\ \bibinfo {pages}
  {1011} (\bibinfo {year} {1992})}\BibitemShut {NoStop}%
\bibitem [{\citenamefont {Bereau}\ and\ \citenamefont
  {Swendsen}(2009)}]{bereau_swendsen09}%
  \BibitemOpen
  \bibfield  {author} {\bibinfo {author} {\bibfnamefont {T.}~\bibnamefont
  {Bereau}}\ and\ \bibinfo {author} {\bibfnamefont {R.~H.}\ \bibnamefont
  {Swendsen}},\ }\href@noop {} {\bibfield  {journal} {\bibinfo  {journal} {J.
  Comput. Phys.}\ }\textbf {\bibinfo {volume} {228}},\ \bibinfo {pages} {6119}
  (\bibinfo {year} {2009})}\BibitemShut {NoStop}%
\bibitem [{\citenamefont {Chernick}(2008)}]{resampling}%
  \BibitemOpen
  \bibfield  {author} {\bibinfo {author} {\bibfnamefont {M.~R.}\ \bibnamefont
  {Chernick}},\ }\href@noop {} {\emph {\bibinfo {title} {Bootstrap Methods: A
  Guide for Practitioners and Researchers}}},\ \bibinfo {edition} {2nd}\ ed.\
  (\bibinfo  {publisher} {Wiley-Interscience},\ \bibinfo {year}
  {2008})\BibitemShut {NoStop}%
\bibitem [{\citenamefont {Hess}\ \emph {et~al.}(2008)\citenamefont {Hess},
  \citenamefont {Kutzner}, \citenamefont {Van Der~Spoel},\ and\ \citenamefont
  {Lindahl}}]{hess2008gromacs}%
  \BibitemOpen
  \bibfield  {author} {\bibinfo {author} {\bibfnamefont {B.}~\bibnamefont
  {Hess}}, \bibinfo {author} {\bibfnamefont {C.}~\bibnamefont {Kutzner}},
  \bibinfo {author} {\bibfnamefont {D.}~\bibnamefont {Van Der~Spoel}}, \ and\
  \bibinfo {author} {\bibfnamefont {E.}~\bibnamefont {Lindahl}},\ }\href@noop
  {} {\bibfield  {journal} {\bibinfo  {journal} {Journal of chemical theory and
  computation}\ }\textbf {\bibinfo {volume} {4}},\ \bibinfo {pages} {435}
  (\bibinfo {year} {2008})}\BibitemShut {NoStop}%
\bibitem [{\citenamefont {Yesylevskyy}\ \emph {et~al.}(2010)\citenamefont
  {Yesylevskyy}, \citenamefont {Sch{\"a}fer}, \citenamefont {Sengupta},\ and\
  \citenamefont {Marrink}}]{yesylevskyy2010polarizable}%
  \BibitemOpen
  \bibfield  {author} {\bibinfo {author} {\bibfnamefont {S.~O.}\ \bibnamefont
  {Yesylevskyy}}, \bibinfo {author} {\bibfnamefont {L.~V.}\ \bibnamefont
  {Sch{\"a}fer}}, \bibinfo {author} {\bibfnamefont {D.}~\bibnamefont
  {Sengupta}}, \ and\ \bibinfo {author} {\bibfnamefont {S.~J.}\ \bibnamefont
  {Marrink}},\ }\href@noop {} {\bibfield  {journal} {\bibinfo  {journal} {PLoS
  Comput Biol}\ }\textbf {\bibinfo {volume} {6}},\ \bibinfo {pages} {e1000810}
  (\bibinfo {year} {2010})}\BibitemShut {NoStop}%
\bibitem [{\citenamefont {Cisneros}\ \emph {et~al.}(2014)\citenamefont
  {Cisneros}, \citenamefont {Karttunen}, \citenamefont {Ren},\ and\
  \citenamefont {Sagui}}]{Cisneros2013}%
  \BibitemOpen
  \bibfield  {author} {\bibinfo {author} {\bibfnamefont {G.~A.}\ \bibnamefont
  {Cisneros}}, \bibinfo {author} {\bibfnamefont {M.}~\bibnamefont {Karttunen}},
  \bibinfo {author} {\bibfnamefont {P.}~\bibnamefont {Ren}}, \ and\ \bibinfo
  {author} {\bibfnamefont {C.}~\bibnamefont {Sagui}},\ }\href {\doibase
  10.1021/cr300461d} {\bibfield  {journal} {\bibinfo  {journal} {Chem Rev}\
  }\textbf {\bibinfo {volume} {114}},\ \bibinfo {pages} {779−814} (\bibinfo
  {year} {2014})}\BibitemShut {NoStop}%
\bibitem [{\citenamefont {Bussi}, \citenamefont {Donadio},\ and\ \citenamefont
  {Parrinello}(2007)}]{bussi2007canonical}%
  \BibitemOpen
  \bibfield  {author} {\bibinfo {author} {\bibfnamefont {G.}~\bibnamefont
  {Bussi}}, \bibinfo {author} {\bibfnamefont {D.}~\bibnamefont {Donadio}}, \
  and\ \bibinfo {author} {\bibfnamefont {M.}~\bibnamefont {Parrinello}},\
  }\href@noop {} {\bibfield  {journal} {\bibinfo  {journal} {The Journal of
  chemical physics}\ }\textbf {\bibinfo {volume} {126}},\ \bibinfo {pages}
  {014101} (\bibinfo {year} {2007})}\BibitemShut {NoStop}%
\bibitem [{\citenamefont {Berendsen}\ \emph {et~al.}(1984)\citenamefont
  {Berendsen}, \citenamefont {Postma}, \citenamefont {van Gunsteren},
  \citenamefont {DiNola},\ and\ \citenamefont {Haak}}]{berendsen1984molecular}%
  \BibitemOpen
  \bibfield  {author} {\bibinfo {author} {\bibfnamefont {H.~J.}\ \bibnamefont
  {Berendsen}}, \bibinfo {author} {\bibfnamefont {J.~P.~M.}\ \bibnamefont
  {Postma}}, \bibinfo {author} {\bibfnamefont {W.~F.}\ \bibnamefont {van
  Gunsteren}}, \bibinfo {author} {\bibfnamefont {A.}~\bibnamefont {DiNola}}, \
  and\ \bibinfo {author} {\bibfnamefont {J.}~\bibnamefont {Haak}},\ }\href@noop
  {} {\bibfield  {journal} {\bibinfo  {journal} {The Journal of chemical
  physics}\ }\textbf {\bibinfo {volume} {81}},\ \bibinfo {pages} {3684}
  (\bibinfo {year} {1984})}\BibitemShut {NoStop}%
\bibitem [{\citenamefont {Hub}, \citenamefont {De~Groot},\ and\ \citenamefont
  {Van Der~Spoel}(2010)}]{hub2010g_wham}%
  \BibitemOpen
  \bibfield  {author} {\bibinfo {author} {\bibfnamefont {J.~S.}\ \bibnamefont
  {Hub}}, \bibinfo {author} {\bibfnamefont {B.~L.}\ \bibnamefont {De~Groot}}, \
  and\ \bibinfo {author} {\bibfnamefont {D.}~\bibnamefont {Van Der~Spoel}},\
  }\href@noop {} {\bibfield  {journal} {\bibinfo  {journal} {Journal of
  Chemical Theory and Computation}\ }\textbf {\bibinfo {volume} {6}},\ \bibinfo
  {pages} {3713} (\bibinfo {year} {2010})}\BibitemShut {NoStop}%
\bibitem [{\citenamefont {Wassenaar}\ \emph {et~al.}(2014)\citenamefont
  {Wassenaar}, \citenamefont {Pluhackova}, \citenamefont {Böckmann},
  \citenamefont {Marrink},\ and\ \citenamefont
  {Tieleman}}]{wassenaar2014going}%
  \BibitemOpen
  \bibfield  {author} {\bibinfo {author} {\bibfnamefont {T.~A.}\ \bibnamefont
  {Wassenaar}}, \bibinfo {author} {\bibfnamefont {K.}~\bibnamefont
  {Pluhackova}}, \bibinfo {author} {\bibfnamefont {R.~A.}\ \bibnamefont
  {Böckmann}}, \bibinfo {author} {\bibfnamefont {S.~J.}\ \bibnamefont
  {Marrink}}, \ and\ \bibinfo {author} {\bibfnamefont {D.~P.}\ \bibnamefont
  {Tieleman}},\ }\href@noop {} {\bibfield  {journal} {\bibinfo  {journal}
  {Journal of Chemical Theory and Computation}\ }\textbf {\bibinfo {volume}
  {10}},\ \bibinfo {pages} {676} (\bibinfo {year} {2014})}\BibitemShut
  {NoStop}%
\bibitem [{\citenamefont {Schmid}\ \emph {et~al.}(2011)\citenamefont {Schmid},
  \citenamefont {Eichenberger}, \citenamefont {Choutko}, \citenamefont
  {Riniker}, \citenamefont {Winger}, \citenamefont {Mark},\ and\ \citenamefont
  {van Gunsteren}}]{schmid2011definition}%
  \BibitemOpen
  \bibfield  {author} {\bibinfo {author} {\bibfnamefont {N.}~\bibnamefont
  {Schmid}}, \bibinfo {author} {\bibfnamefont {A.~P.}\ \bibnamefont
  {Eichenberger}}, \bibinfo {author} {\bibfnamefont {A.}~\bibnamefont
  {Choutko}}, \bibinfo {author} {\bibfnamefont {S.}~\bibnamefont {Riniker}},
  \bibinfo {author} {\bibfnamefont {M.}~\bibnamefont {Winger}}, \bibinfo
  {author} {\bibfnamefont {A.~E.}\ \bibnamefont {Mark}}, \ and\ \bibinfo
  {author} {\bibfnamefont {W.~F.}\ \bibnamefont {van Gunsteren}},\ }\href@noop
  {} {\bibfield  {journal} {\bibinfo  {journal} {European biophysics journal}\
  }\textbf {\bibinfo {volume} {40}},\ \bibinfo {pages} {843} (\bibinfo {year}
  {2011})}\BibitemShut {NoStop}%
\bibitem [{\citenamefont {Poger}, \citenamefont {Van~Gunsteren},\ and\
  \citenamefont {Mark}(2010)}]{poger2010new}%
  \BibitemOpen
  \bibfield  {author} {\bibinfo {author} {\bibfnamefont {D.}~\bibnamefont
  {Poger}}, \bibinfo {author} {\bibfnamefont {W.~F.}\ \bibnamefont
  {Van~Gunsteren}}, \ and\ \bibinfo {author} {\bibfnamefont {A.~E.}\
  \bibnamefont {Mark}},\ }\href@noop {} {\bibfield  {journal} {\bibinfo
  {journal} {Journal of computational chemistry}\ }\textbf {\bibinfo {volume}
  {31}},\ \bibinfo {pages} {1117} (\bibinfo {year} {2010})}\BibitemShut
  {NoStop}%
\bibitem [{\citenamefont {Berendsen}\ \emph {et~al.}(1981)\citenamefont
  {Berendsen}, \citenamefont {Postma}, \citenamefont {van Gunsteren},\ and\
  \citenamefont {Hermans}}]{berendsen1981interaction}%
  \BibitemOpen
  \bibfield  {author} {\bibinfo {author} {\bibfnamefont {H.~J.}\ \bibnamefont
  {Berendsen}}, \bibinfo {author} {\bibfnamefont {J.~P.}\ \bibnamefont
  {Postma}}, \bibinfo {author} {\bibfnamefont {W.~F.}\ \bibnamefont {van
  Gunsteren}}, \ and\ \bibinfo {author} {\bibfnamefont {J.}~\bibnamefont
  {Hermans}},\ }in\ \href@noop {} {\emph {\bibinfo {booktitle} {Intermolecular
  forces}}}\ (\bibinfo  {publisher} {Springer},\ \bibinfo {year} {1981})\ pp.\
  \bibinfo {pages} {331--342}\BibitemShut {NoStop}%
\bibitem [{\citenamefont {Hess}\ \emph {et~al.}(1997)\citenamefont {Hess},
  \citenamefont {Bekker}, \citenamefont {Berendsen}, \citenamefont {Fraaije}
  \emph {et~al.}}]{hess1997lincs}%
  \BibitemOpen
  \bibfield  {author} {\bibinfo {author} {\bibfnamefont {B.}~\bibnamefont
  {Hess}}, \bibinfo {author} {\bibfnamefont {H.}~\bibnamefont {Bekker}},
  \bibinfo {author} {\bibfnamefont {H.~J.}\ \bibnamefont {Berendsen}}, \bibinfo
  {author} {\bibfnamefont {J.~G.}\ \bibnamefont {Fraaije}},  \emph {et~al.},\
  }\href@noop {} {\bibfield  {journal} {\bibinfo  {journal} {Journal of
  computational chemistry}\ }\textbf {\bibinfo {volume} {18}},\ \bibinfo
  {pages} {1463} (\bibinfo {year} {1997})}\BibitemShut {NoStop}%
\bibitem [{\citenamefont {Essmann}\ \emph {et~al.}(1995)\citenamefont
  {Essmann}, \citenamefont {Perera}, \citenamefont {Berkowitz}, \citenamefont
  {Darden}, \citenamefont {Lee},\ and\ \citenamefont
  {Pedersen}}]{essmann1995smooth}%
  \BibitemOpen
  \bibfield  {author} {\bibinfo {author} {\bibfnamefont {U.}~\bibnamefont
  {Essmann}}, \bibinfo {author} {\bibfnamefont {L.}~\bibnamefont {Perera}},
  \bibinfo {author} {\bibfnamefont {M.~L.}\ \bibnamefont {Berkowitz}}, \bibinfo
  {author} {\bibfnamefont {T.}~\bibnamefont {Darden}}, \bibinfo {author}
  {\bibfnamefont {H.}~\bibnamefont {Lee}}, \ and\ \bibinfo {author}
  {\bibfnamefont {L.~G.}\ \bibnamefont {Pedersen}},\ }\href@noop {} {\bibfield
  {journal} {\bibinfo  {journal} {The Journal of chemical physics}\ }\textbf
  {\bibinfo {volume} {103}},\ \bibinfo {pages} {8577} (\bibinfo {year}
  {1995})}\BibitemShut {NoStop}%
\bibitem [{\citenamefont {Bussi}\ \emph {et~al.}(2006)\citenamefont {Bussi},
  \citenamefont {Gervasio}, \citenamefont {Laio},\ and\ \citenamefont
  {Parrinello}}]{bussi2006free}%
  \BibitemOpen
  \bibfield  {author} {\bibinfo {author} {\bibfnamefont {G.}~\bibnamefont
  {Bussi}}, \bibinfo {author} {\bibfnamefont {F.~L.}\ \bibnamefont {Gervasio}},
  \bibinfo {author} {\bibfnamefont {A.}~\bibnamefont {Laio}}, \ and\ \bibinfo
  {author} {\bibfnamefont {M.}~\bibnamefont {Parrinello}},\ }\href@noop {}
  {\bibfield  {journal} {\bibinfo  {journal} {Journal of the American Chemical
  Society}\ }\textbf {\bibinfo {volume} {128}},\ \bibinfo {pages} {13435}
  (\bibinfo {year} {2006})}\BibitemShut {NoStop}%
\bibitem [{\citenamefont {Bonomi}\ and\ \citenamefont
  {Parrinello}(2010)}]{bonomi2010enhanced}%
  \BibitemOpen
  \bibfield  {author} {\bibinfo {author} {\bibfnamefont {M.}~\bibnamefont
  {Bonomi}}\ and\ \bibinfo {author} {\bibfnamefont {M.}~\bibnamefont
  {Parrinello}},\ }\href@noop {} {\bibfield  {journal} {\bibinfo  {journal}
  {Physical review letters}\ }\textbf {\bibinfo {volume} {104}},\ \bibinfo
  {pages} {190601} (\bibinfo {year} {2010})}\BibitemShut {NoStop}%
\bibitem [{\citenamefont {Deighan}, \citenamefont {Bonomi},\ and\ \citenamefont
  {Pfaendtner}(2012)}]{deighan2012efficient}%
  \BibitemOpen
  \bibfield  {author} {\bibinfo {author} {\bibfnamefont {M.}~\bibnamefont
  {Deighan}}, \bibinfo {author} {\bibfnamefont {M.}~\bibnamefont {Bonomi}}, \
  and\ \bibinfo {author} {\bibfnamefont {J.}~\bibnamefont {Pfaendtner}},\
  }\href@noop {} {\bibfield  {journal} {\bibinfo  {journal} {Journal of
  Chemical Theory and Computation}\ }\textbf {\bibinfo {volume} {8}},\ \bibinfo
  {pages} {2189} (\bibinfo {year} {2012})}\BibitemShut {NoStop}%
\bibitem [{\citenamefont {Tribello}\ \emph {et~al.}(2014)\citenamefont
  {Tribello}, \citenamefont {Bonomi}, \citenamefont {Branduardi}, \citenamefont
  {Camilloni},\ and\ \citenamefont {Bussi}}]{tribello2014plumed}%
  \BibitemOpen
  \bibfield  {author} {\bibinfo {author} {\bibfnamefont {G.~A.}\ \bibnamefont
  {Tribello}}, \bibinfo {author} {\bibfnamefont {M.}~\bibnamefont {Bonomi}},
  \bibinfo {author} {\bibfnamefont {D.}~\bibnamefont {Branduardi}}, \bibinfo
  {author} {\bibfnamefont {C.}~\bibnamefont {Camilloni}}, \ and\ \bibinfo
  {author} {\bibfnamefont {G.}~\bibnamefont {Bussi}},\ }\href@noop {}
  {\bibfield  {journal} {\bibinfo  {journal} {Computer Physics Communications}\
  }\textbf {\bibinfo {volume} {185}},\ \bibinfo {pages} {604} (\bibinfo {year}
  {2014})}\BibitemShut {NoStop}%
\bibitem [{\citenamefont {Do}, \citenamefont {Choy},\ and\ \citenamefont
  {Karttunen}(2014)}]{Do2014}%
  \BibitemOpen
  \bibfield  {author} {\bibinfo {author} {\bibfnamefont {T.~N.}\ \bibnamefont
  {Do}}, \bibinfo {author} {\bibfnamefont {W.-Y.}\ \bibnamefont {Choy}}, \ and\
  \bibinfo {author} {\bibfnamefont {M.}~\bibnamefont {Karttunen}},\ }\href@noop
  {} {\bibfield  {journal} {\bibinfo  {journal} {Journal of Chemical Theory and
  Computation}\ }\textbf {\bibinfo {volume} {10}},\ \bibinfo {pages} {5081}
  (\bibinfo {year} {2014})}\BibitemShut {NoStop}%
\bibitem [{\citenamefont {Barducci}, \citenamefont {Bussi},\ and\ \citenamefont
  {Parrinello}(2008)}]{barducci2008well}%
  \BibitemOpen
  \bibfield  {author} {\bibinfo {author} {\bibfnamefont {A.}~\bibnamefont
  {Barducci}}, \bibinfo {author} {\bibfnamefont {G.}~\bibnamefont {Bussi}}, \
  and\ \bibinfo {author} {\bibfnamefont {M.}~\bibnamefont {Parrinello}},\
  }\href@noop {} {\bibfield  {journal} {\bibinfo  {journal} {Physical review
  letters}\ }\textbf {\bibinfo {volume} {100}},\ \bibinfo {pages} {020603}
  (\bibinfo {year} {2008})}\BibitemShut {NoStop}%
\bibitem [{\citenamefont {Bereau}(2011)}]{bereau2011unconstrained}%
  \BibitemOpen
  \bibfield  {author} {\bibinfo {author} {\bibfnamefont {T.}~\bibnamefont
  {Bereau}},\ }\emph {\bibinfo {title} {Unconstrained Structure Formation in
  Coarse-Grained Protein Simulations}},\ \href@noop {} {Ph.D. thesis},\
  \bibinfo  {school} {Carnegie Mellon University} (\bibinfo {year}
  {2011})\BibitemShut {NoStop}%
\bibitem [{\citenamefont {Shirts}\ and\ \citenamefont
  {Chodera}(2008)}]{shirts2008statistically}%
  \BibitemOpen
  \bibfield  {author} {\bibinfo {author} {\bibfnamefont {M.~R.}\ \bibnamefont
  {Shirts}}\ and\ \bibinfo {author} {\bibfnamefont {J.~D.}\ \bibnamefont
  {Chodera}},\ }\href@noop {} {\bibfield  {journal} {\bibinfo  {journal} {The
  Journal of chemical physics}\ }\textbf {\bibinfo {volume} {129}},\ \bibinfo
  {pages} {124105} (\bibinfo {year} {2008})}\BibitemShut {NoStop}%
\bibitem [{\citenamefont {Zhu}\ and\ \citenamefont
  {Hummer}(2012)}]{zhu2012convergence}%
  \BibitemOpen
  \bibfield  {author} {\bibinfo {author} {\bibfnamefont {F.}~\bibnamefont
  {Zhu}}\ and\ \bibinfo {author} {\bibfnamefont {G.}~\bibnamefont {Hummer}},\
  }\href@noop {} {\bibfield  {journal} {\bibinfo  {journal} {Journal of
  computational chemistry}\ }\textbf {\bibinfo {volume} {33}},\ \bibinfo
  {pages} {453} (\bibinfo {year} {2012})}\BibitemShut {NoStop}%
\bibitem [{\citenamefont {Rawicz}\ \emph {et~al.}(2000)\citenamefont {Rawicz},
  \citenamefont {Olbrich}, \citenamefont {McIntosh}, \citenamefont {Needham},\
  and\ \citenamefont {Evans}}]{rawicz2000effect}%
  \BibitemOpen
  \bibfield  {author} {\bibinfo {author} {\bibfnamefont {W.}~\bibnamefont
  {Rawicz}}, \bibinfo {author} {\bibfnamefont {K.}~\bibnamefont {Olbrich}},
  \bibinfo {author} {\bibfnamefont {T.}~\bibnamefont {McIntosh}}, \bibinfo
  {author} {\bibfnamefont {D.}~\bibnamefont {Needham}}, \ and\ \bibinfo
  {author} {\bibfnamefont {E.}~\bibnamefont {Evans}},\ }\href@noop {}
  {\bibfield  {journal} {\bibinfo  {journal} {Biophysical journal}\ }\textbf
  {\bibinfo {volume} {79}},\ \bibinfo {pages} {328} (\bibinfo {year}
  {2000})}\BibitemShut {NoStop}%
\end{thebibliography}%

\end{document}